\documentclass[journal]{IEEEtran}
\usepackage{xcolor,soul,framed} %,caption
\colorlet{shadecolor}{yellow}
\usepackage[pdftex]{graphicx}
\usepackage{caption}
\usepackage{subcaption}
\graphicspath{{../pdf/}{../jpeg/}}
\DeclareGraphicsExtensions{.pdf,.jpeg,.png}
\usepackage[cmex10]{amsmath}
\usepackage{amssymb}
\usepackage{amsthm}
\usepackage{amsmath}
\usepackage{array}
\usepackage{mdwmath}
\usepackage{mdwtab}
\usepackage{eqparbox}
\usepackage{url}
\usepackage{tabularx}
\usepackage{booktabs}
\usepackage{algorithm}
\usepackage{algorithmic}
\usepackage{inputenc}
\usepackage{graphicx}
\newtheorem*{remark}{Remark}
\begin{document}
\bstctlcite{IEEEexample:BSTcontrol}
    % \title{Beam Selection and Reforming for Hovering-Tolerant UAVs Communication over Interference Channels: A Deep-Q Learning for Intelligent Space-Air-Ground Integrated Networks}

    % \title{Distributed 3D-Beam Reforming for Hovering-Tolerant UAVs Communication over Interference: A Deep-Q Learning for Intelligent Space-Air-Ground Integrated Networks}

    \title{Distributed 3D-Beam Reforming for Hovering-Tolerant UAVs Communication over Coexistence: A Deep-Q Learning for Intelligent Space-Air-Ground Integrated Networks}
  \author{Sudhanshu~Arya,~\IEEEmembership{Member,~IEEE},~
          Yifeng Peng,~\IEEEmembership{Student Member,~IEEE},~ 
          Jingda Yang,~\IEEEmembership{Student Member,~IEEE},~         
      and~Ying~Wang,~\IEEEmembership{Member,~IEEE}% <-this % stops a space

  \thanks{(Corresponding author: Ying Wang.)}
  \thanks{Sudhanshu~Arya, Yifeng~Peng, Jingda Yang, and Ying~Wang are with the School of Systems and Enterprises, Stevens Institute of Technology, Hoboken, USA (e-mail: sarya@stevens.edu; ypeng21@stevens.edu; jyang76@stevens.edu; ywang6@stevens.edu).}% <-this % stops a space
  }

% The paper headers
\markboth{}{Arya \MakeLowercase{\textit{et al.}}: UAV}

% ====================================================================
\maketitle

\begin{abstract}
Unmanned aerial vehicles (UAVs) enabled wireless communication is envisioned as a promising technology to realize long transmission distances with seamless coverage for next-generation wireless systems and serving as aerial access points in  Non-Terrestrial Network (NTN) and Space-Air-Ground Integrated Networks (SAGIN), especially for Sub-6 GHz spectrum range. In this paper, we present a novel distributed UAVs beam reforming approach to dynamically form and reform a space-selective beam path in addressing the coexistence with satellite and terrestrial communications. Despite the unique advantage in spectrum efficiency and security of UAVs performed beamforming, the challenges reside in the array responses' sensitivity to random rotational motion and the hovering nature of the UAVs. This random fluctuation leads to the antenna gain mismatch between the transmitting UAVs and the receiver. Due to this random fluctuation, the target coverage area of the UAVs link changes frequently and thereby requires a hovering-tolerant flexible beamforming technique. In this paper, a model-free reinforcement learning (RL) based unified UAV beam selection and tracking approach is presented to effectively realize the dynamic distributed and collaborative beamforming. The combined impact of the UAVs’ hovering and rotational motions is considered while addressing the impairment due to the interference from the orbiting satellites and neighboring networks. The main objectives of this work are two-fold: first, to acquire the channel awareness between the UAVs and user equipment (UE) to uncover the impairments due to space interference, and to identify the availability of the best links for the provision of opportunistic access to avoid interference; second, to overcome the beam distortion by configuring the selected UAVs for beam re-forming to meet the quality of service (QoS) requirements. In particular, we study the variations in the angle of arrival and show that the beam does not accurately point toward the receiver due to the random hovering and rotational motion of the UAVs. To overcome the impact of the interference and to maximize the beamforming gain with minimum squint losses, we define and apply a new optimal UAVs selection algorithm based on the brute force criteria. The proposed selection algorithm allows the detection of the best optimal UAVs for beamforming, based on their orientations and channel conditions. Results demonstrate that the detrimental effects of the channel fading and the interference from the orbiting satellites and from the neighboring network can be overcome using the proposed approach. Subsequently, an RL algorithm based on Deep Q-Network (DQN) is developed for real-time beam tracking. A Deep Neural Network (DNN) is utilized as an approximator function to estimate the Q-values. We consider the idea of experience replay to enable the DQN agent to learn from experience. By augmenting the system with the impairments due to the hovering and rotational motion, we show that the proposed DQN algorithm can reform the beam in real time with negligible mean square error (MSE) without adding any additional cost. It is demonstrated that the beam re-forming technique based on the proposed DQN algorithm attains an exceptional performance improvement. Moreover, we present the results which show fast convergence of the proposed DQN algorithm. We show that when the number of best UAVs required is set to $4$ for a 3D UAVs network with $64$ UAVs arranged in a rectangular geometry, the proposed approach requires approximately 50 iterations only while fine-tuning its parameters quickly without observing any plateaus. Importantly, we also show that the learning algorithm performs efficiently irrespective of the hovering tolerance value.

\end{abstract}


\IEEEpeerreviewmaketitle

\section{Introduction}

\IEEEPARstart{W}{ith} the inclusion of Non-Terrestrial Network (NTN) and Satellite work item in 3rd Generation Partnership Project (3GPP) fifth-generation (5G) Release 17 and Release 18, satellites communications have been considered as an effective complementary extension of the 5G terrestrial networks coverage, especially over low density populated areas \cite{Lin20215GNetworks}. One of the challenges is to ensure the quality and cost of the NTN services are comparable to the terrestrial counterpart, particularly in the below 6 GHz satellite bands.  With the stringent power flux density limitations and spectrum sparsity in both ground and satellite communications, the order of magnitude effective delivered throughput increase and service cost reduction demands require an innovative approach in connecting the space and ground segment. The expected large-scale deployment of Satellite Communication in the near future, including widespread Starlink applications requires maximizing the utilization of space in SAGIN with the significantly increasing communication capacity and coexisting with existing Satellite communication.

With the rapid progress in developing 5G and Beyond wireless communication systems and technologies have entered the stage of vast commercial and industrial deployment with improved end-user experiences. The scalable, flexible yet high-performance payloads architectures and technologies are crucial to cope with 5G and beyond requirements.  In context to the limitations in reliability, resilience, and performance in 5G and beyond , the key open challenges worthy of further research are enabling wider coverage with ultra-dense connection, ultra-low latency, and ubiquitous intelligence \cite{wang2023road, geraci2022will, de2021survey}. Among the enabling energy-efficient technologies, beamforming-enabled Massive MIMO provides benefits in terms of power gains, which could be translated into dramatic increases in range, rate, or energy efficiency, as well as security and interference reduction since less transmit power is scattered in unintended directions\cite{Bjornson2019MassiveArrays}\cite{Mudumbai2009DistributedProgress}. Meanwhile, the application and deployment of beam forming are challenged by the stringent requirements in hardware and computation complexity for carrying fixed-distance antenna arrays. Distributed transmit beamforming, a form of cooperative communication in which two or more information sources simultaneously transmit a common message and control the phase of their transmissions to  constructively combine signals at an intended destination has shown promising aspects, especially in ad-hoc networks, over the past decades \cite{Mudumbai2009DistributedProgress}.

Wireless communications involving unmanned aerial vehicles (UAVs) have seen a surge of interest due to their capabilities to enable wider coverage. UAVs are expected to play a major role in next-generation cellular communications by providing wider coverage with ubiquitous connectivity \cite{polus2023performance} and serving as aerial access points in NTN and Space-Air-Ground Integrated Networks (SAGIN) in Sub-6 GHz \cite{Lin20215GNetworks}. The Distributed\/Collaborative beamforming of UAVs provides unique advantages in communication performance, seamless coverage, resilience to interference and attack, hardware affordability, flexibility in beam shaping, and dynamic adaptation to the environment, however, sensitive to random rotational motion and hovering nature of the UAVs.

\subsection{Preliminary Studies}

To realize flexible and wider coverage for mmWave-based UAV-to-ground communications, a low-complex 3D beamforming approach was proposed by utilizing a uniform planar array (UPA) equipped on the single UAV \cite{zhu20193}. This wider beamforming approach was based on the concept of transforming the target coverage area into the special angles coordinates and then selecting the sub-arrays of the UPA to steer the beam. Though this 3D beamforming approach was considered suitable for mobile scenarios and can be adjusted dynamically based on the UAV and ground-based receiver locations, no synchronization algorithm or mechanism was provided or discussed in selecting the optimal sub-array. Moreover, from the design perspective, it was assumed that the number of antennas equipped on the UAV is 64 $\times$ 64 which requires careful engineering imposed with multiple constraints or challenges such as power requirements, size and weight constraints and increased mutual coupling in the case of mmWave.

In another similar study, a single UAV equipped with a large-scale UPA was considered to communicate with multiple ground users via mmWave links \cite{liu2023deployment}. To account for the impairment in the communication quality, the impact of UAV jitter was considered. In particular, a model was developed to analyze the relationship between the UAV jitter and the deviation in the angle of departure. A non-convex problem was formulated to maximize the minimum achievable throughput for all the ground-located users. To analyze the performance, the authors considered 16 $\times$ 16, 32 $\times$ 32, 64 $\times$ 64, and 128 $\times$ 128 UPA sizes. Though the findings of this study made important contributions, it is important to note that, the analysis presented considered only the impact of the UAV jitter while neglecting the location uncertainty and displacement of the UAV due to hovering.

In another study, a UAV-enabled virtual antenna array was considered to construct a collaborative beam to transmit the information toward the ground-located cluster of base stations \cite{sun2023uav}. In this approach, the authors assumed multiple known and unknown eavesdroppers aiming to listen to the transmitted information and therefore considered the security analysis. To achieve security, a salp swarm intelligent algorithm-based non-convex and NP-hard multi-objective optimization problem was formulated to obtain the optimized positions of the UAV. The objective of this study was to enable secure communication and achieve the maximum worst-case secrecy rate. However, we like to point out that, in this study, the impact of the UAVs hovering was ignored. Moreover, performance impairments due to timing and phase synchronization errors were also neglected.

Recently, machine learning has been seen as an important enabler for 5G and next-generation cellular communication technologies. It envisioned to provide solutions to many critical challenges including admission control for network slicing, optimization of massive multiple-input multiple-output (MIMO), identification and authentication of the user equipment (UE)\cite{yang20235g}, identify potential interference and jamming\cite{wang2021ai}\cite{Wang2022AnonymousShetty}, tracking and estimation of the propagation channel\cite{mak2023characterization}, location\cite{Wang2021DevelopmentResearch}, and dynamic spectrum sharing \cite{sagduyu2022task, arya2022novel, ayed2023accordion}. Moreover, considering UAV swarms over a flying Ad-Hoc network, a deep reinforcement learning algorithm was presented to solve the decomposition problem of large-scale UAV swarms \cite{zhang2023decomposing}.

Even when the information on the underlying UAV channel characteristics and interference from the satellite and neighboring networks are known, this information may change rapidly with time due to the dynamic channel conditions along with the uncertainty in the neighboring networks and interference from the orbiting satellite. Therefore, the conventional model-based detection methods that rely on the instantaneous estimation of the channel quality, can make a good choice for dynamic UAV channels. However, these methods entail overhead that in turn increase the latency or reduce the information transfer rate.  In addition, the accuracy of these methods may also greatly impact the overall performance \cite{yang2022channel}.

In addition, as an inescapable obstacle, interference from the adjacent networks and orbiting satellites unavoidably leads to the performance degradation of the ground-located BS-to-air and air-to-air (A2A) communications. Moreover, sharing of common channels or resources by satellites, aerials, and terrestrial sources results in potentially serious interference problems \cite{weerackody2016sensitivity, euler2023using}. These interference sources are difficult to determine due to their wide range of potential locations. The satellites and the neighboring interference sources generate a random interference signal around the underlying network, thereby interfering with the information signal \cite{ayoubi2023imt}.

A secure and energy-efficient UAV-enabled relay communication was documented \cite{sun2022secure}. To further enhance the signal strength and performance, the authors considered a collaborative beamforming technique by utilizing a virtual antenna array. A multi-objective optimization problem was formulated to maximize the total minimum secrecy rates while minimizing the sidelobes of the virtual antenna array. In another work, a multi-objective swarm intelligence-based collaborative beamforming technique was proposed using a virtual antenna array mounted on the UAV \cite{sun2022collaborative}. The authors optimized the UAVs' positions and beamforming weights while maximizing the secrecy rates. 

In this paper, we consider a ground-to-air wireless communication scenario where ground situated base station (BS) communicates with the user equipment (UE) via a network of UAVs. As illustrated in Fig. \ref{FIG-1}, the number of distributed UAVs in the network are selected to form a directive beam towards the UE. Depending on the instantaneous location of the UAVs, the optimal UAVs are selected from the network to circumvent the blockage and interference signals from the neighboring networks and from the orbiting satellites. It should be noted that, depending on the UAVs’ locations, the channel between the distributed UAVs and UE could either experience huge interference or poor propagation conditions. Therefore, the most appropriate UAVs should be selected to construct beamforming for optimal performance. To this end, we formulate a combinatorial optimization problem for the best subset selection, such that, the best combination of $K$ UAVs can be selected out of $N$ UAVs, $K \leq N$. Since it requires evaluating all possible combinations of $K$ UAVs, the proposed algorithm is an exhaustive search algorithm based on the brute-force approach. However, due to the hovering and mobility of selected UAVs, it leads to frequent beam misalignment. Motivated by this, we further proposed a Deep Q-Learning-based reinforcement learning algorithm for beam reforming. The proposed DQN returns real-time optimal location coordinates for the selected UAVs to the control center to avoid beam distortion due to misalignment.

\subsection{Motivations}
Although UAVs-enabled aerial access points in cellular communications offer a significant improvement in achieving wider coverage, it still has several limitations. In UAVs' cellular communications, the ground base stations are typically down-tilted. This results in a sharp signal fluctuation at the UAVs since the UAVs can only be reached by the upper antenna side lobes of the base station. Moreover, UAVs receive and transmit interfering signals from/to a plurality of neighboring networks, cells, and signals from the satellites, and therefore hinder the correct decoding of the information signals \cite{geraci2022will}. Moreover, rotational motion and hovering inaccuracy of the UAVs introduce localization and orientation mismatch. Hovering inaccuracy results in a significant degradation of the communication performance due to the reduced received power at UE \cite{suman2020optimal}.

\subsubsection{Limitations of Channel Prediction in a Jitter-and-Interference-Dominated System}
\begin{figure*}
    \centering
\includegraphics[width=\textwidth]{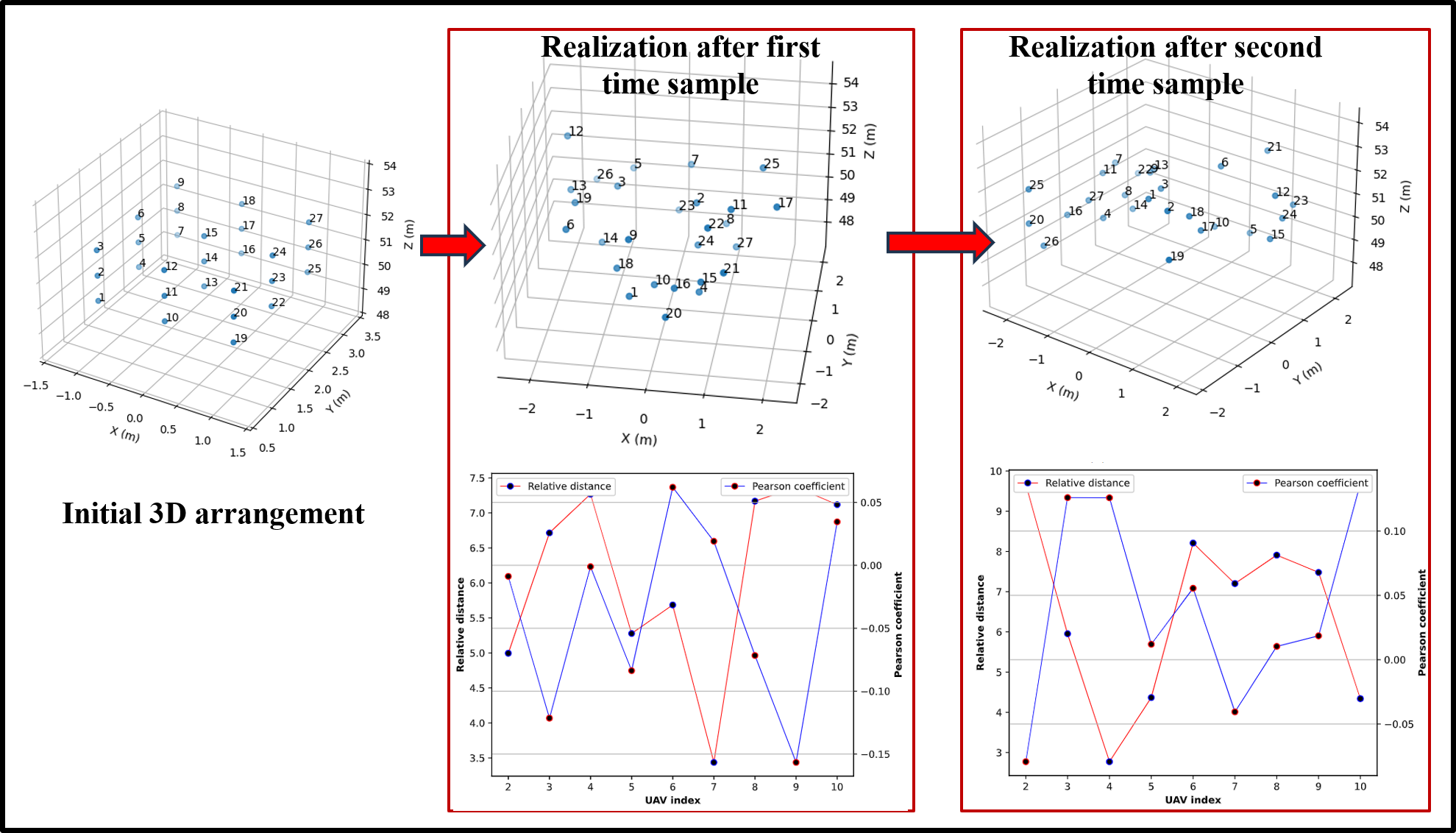}
\caption{Relative distances (in meters) of UAV indexed '1' with other UAV nodes and the corresponding Pearson correlation coefficients over different time samples. \label{FIG-2}}
\end{figure*}

It is important to note that obtaining accurate channel quality information is crucial for effective beamforming. Wireless communication systems operate in a dynamic environment where the surrounding channel conditions can change over time. To characterize the behavior of the dynamic link and to understand the implications of the random walk due to hovering, we harness the knowledge of the Pearson correlation coefficient, a statistical measure to quantify the strength and direction of the relationship between the relative UAVs' distances and the channel fading. Interestingly, the random UAV trajectories due to the hovering and the fast dynamic channel conditions, result in a \textbf{no clear} pattern between the relative UAVs' distances and the fading correlation. We consider two channel realizations corresponding to different time samples to illustrate the channel dynamics over hovering conditions. As illustrated in Fig. \ref{FIG-2}, positive values of the Pearson coefficient indicate the positive relationship between the relative distance and the fading correlation. However, in contrast, there are a few scenarios for which the coefficient does not reach the positive value indicating the negative relation between the relative distance and channel fading.

\begin{figure*}
    \centering
\includegraphics[width=\textwidth]{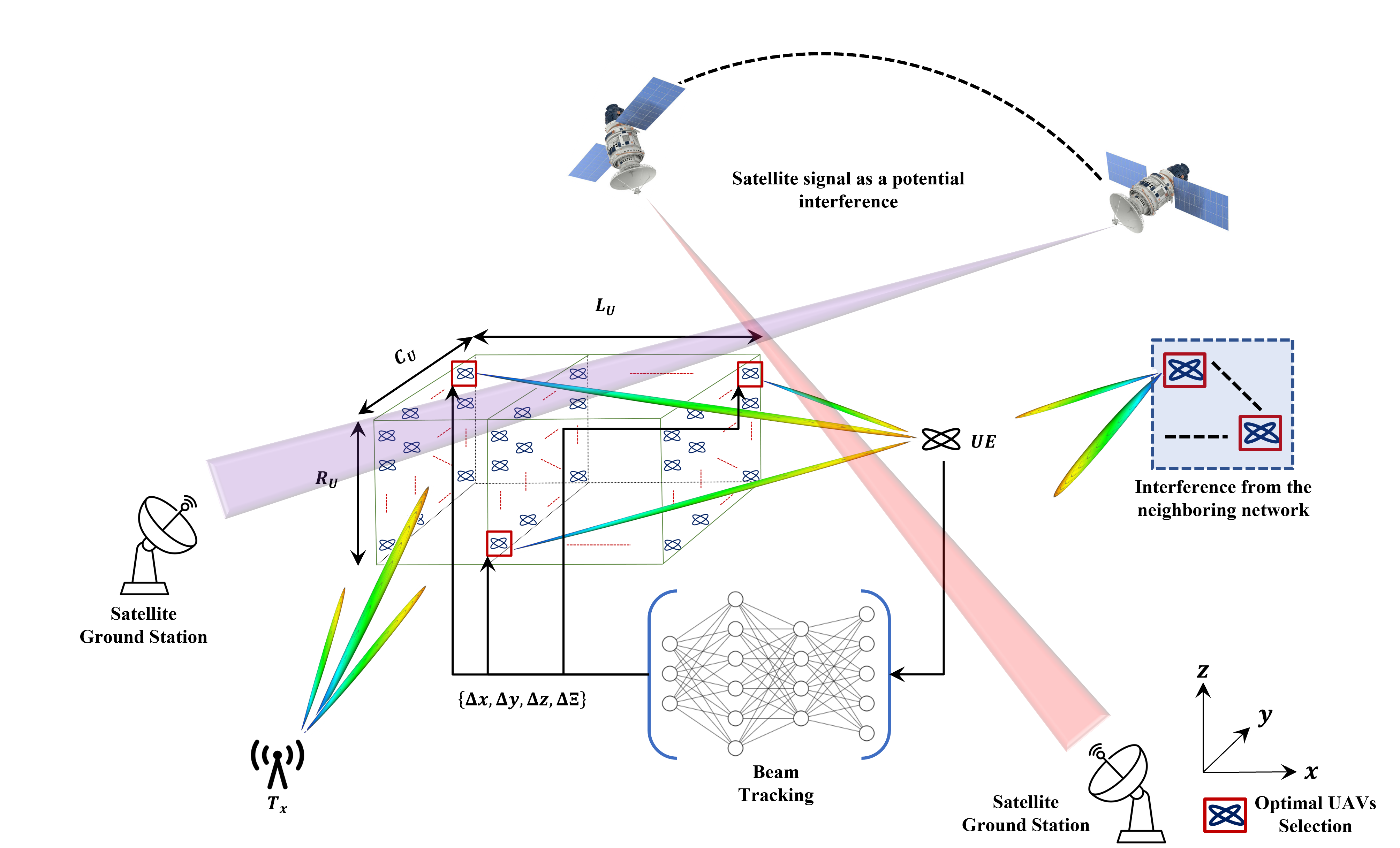}
\caption{Illustration of the system model.\label{FIG-1}}
\end{figure*}

The results presented in Fig. \ref{FIG-2}  underline the limitations of channel prediction in a jitter-and-interference-dominated UAV system.

Moreover, the interference signal arriving from directions other than the desired signal direction can distort the main beam and thereby can cause beam broadening or sidelobe leakage. The interference caused due to satellite signal leakage, overlapping frequency bands, or due to the moving object, can change in time both in terms of signal strength and direction. Moreover, the time-varying channel can result in fluctuations in the interference levels. Therefore, it is important to optimize the UAV selection and orientation to avoid beam distortion due to interference. The regions with potentially high interference levels should be avoided. Figure \ref{FIG-2AA} represents the distribution and intensity of the interference signals at different locations for two different scenarios. The \textbf{2D square plane} represents the region with a strong interference power level.

\subsubsection{Increased Sensitivity of Array Responses to Angle of Arrival(AoA)}
The array response to Angle of Arrival (AoA) in beamforming is impacted by four factors: frequency, spacing between antennas, the location stability of the signal source, and the number of antennas. Compared to fixed-antenna array beamforming operating at mmWave frequencies on the ground, our proposed solution emulates antenna arrays on UAVs, operating at sub-6G (C-band, 3.5 GHz) with longer distances between each antenna due to the safety requirements for UAVs. The signals are transmitted from less stable sources due to the fluctuation and hovering nature of UAVs. All three factors negatively impact the distributed UAV beamforming system, leading to increased sensitivity of the array response, resulting in performance degradation on average and instability. However, the increased sensitivity of array responses improves the network's ability to focus more energy in a specific direction and effectively distinguish signals arriving from different directions at the receiver. Therefore, distributed UAV beamforming at sub-6G can significantly improve the system capacity and interference resilience, provided the challenges associated with UAV movement fluctuations and hovering during communication are addressed.

\subsubsection{Interference to/from the Neighboring Networks}
The beamforming has the potential to cause interference with the neighboring networks or to the signal from the satellites. The following scenarios may occur:
\begin{itemize}
    \item If the beamforming antenna array is misaligned or not carefully calibrated, the focused beam may deviate from the intended direction and can cause interference to the neighboring network.
    \item Moreover, if the side lobes overlap with the sensitive receiver of the neighboring network, it may cause interference.
\end{itemize}

\begin{figure}
    \centering
\includegraphics[width=10.0 cm]{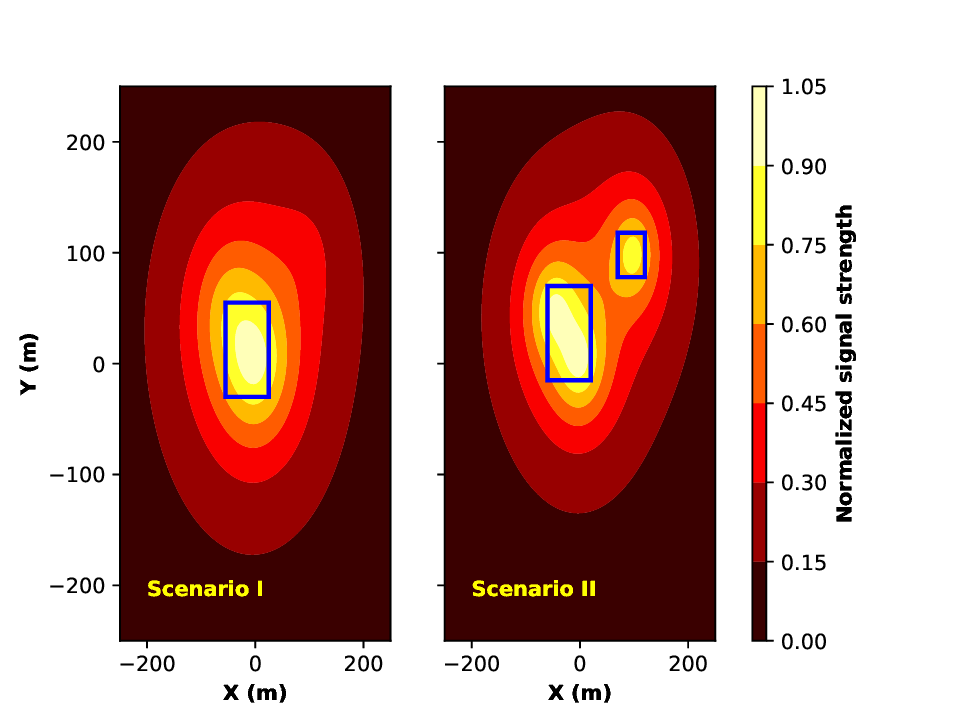}
\caption{Interference heat map representing the distribution and intensity of the interference signals for two different scenarios.\label{FIG-2AA}}
\end{figure}

\subsection{Contributions}
The main contributions of this paper are listed below.
\begin{itemize}
    \item The proposed learning algorithm is computationally efficient and is capable of learning and representing complex and non-linear relationships between selecting the optimal UAVs, beamforming, and beam-retracing. It leads to better generalization and learning capacity. Moreover, it helps alleviate the vanishing gradient problem that can occur in deep neural networks. Importantly, it converges faster due to its non-saturating nature and does not saturate to very low or high values as the input signal variations decrease or increase. Importantly, we also show that the learning algorithm performs efficiently irrespective of the hovering tolerance value.
\end{itemize}

\section{System Model}
\subsection{Impact of Hovering}

\begin{figure}
    \centering
\includegraphics[width=9.0 cm]{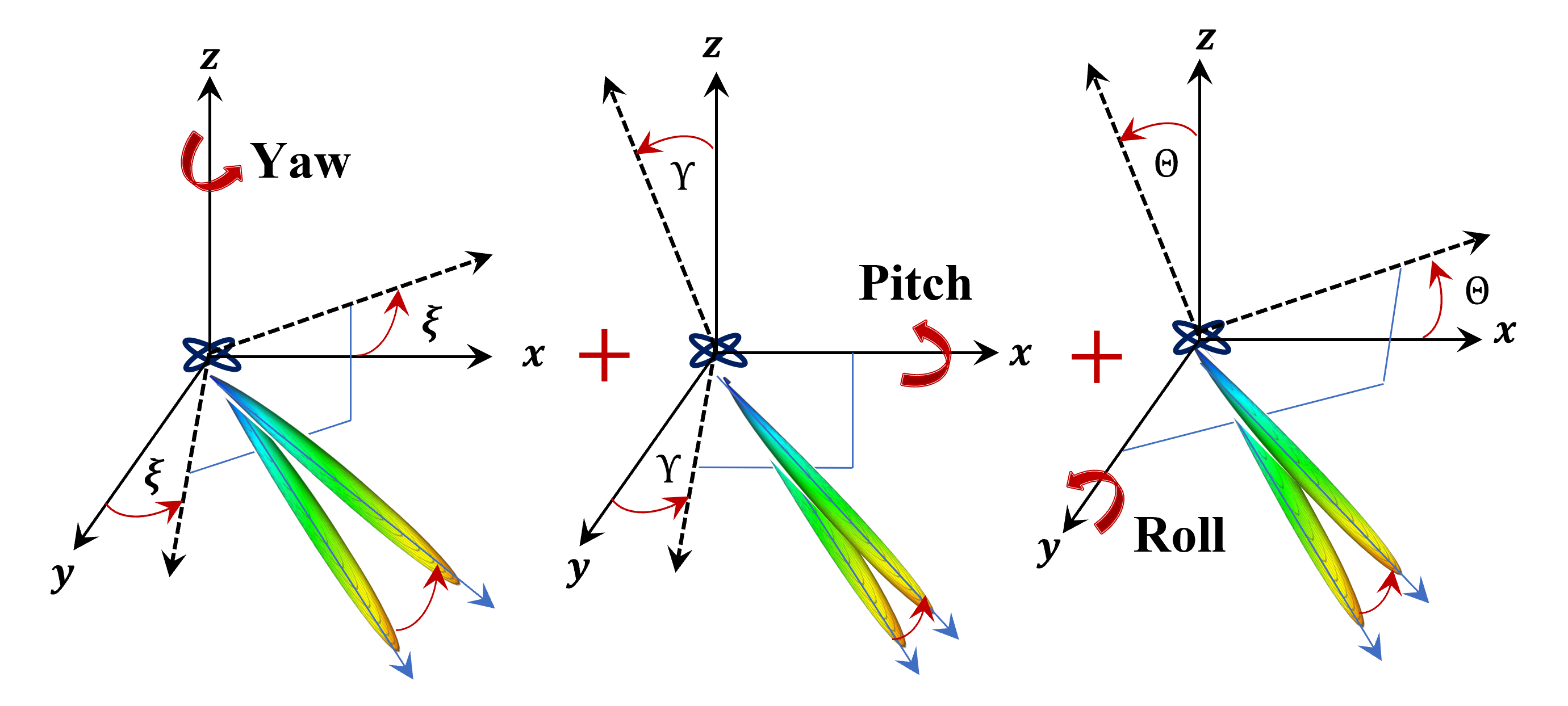}
\caption{Jitter noise due to hovering: Categorization of the rotational motion of the UAV.\label{FIG-2AAA}}
\end{figure}

As illustrated in Fig. \ref{FIG-2AAA}, the rotational motion of the UAV can be categorized into three types: yaw $\xi$, pitch $\gamma$, and roll $\Theta$. Yaw corresponds to the rotation motion around the vertical plane, pitch represents the rotation motion around the wings of the lateral axis, whereas roll is the rotation motion around the head or the longitudinal axis.

\begin{figure}
    \centering
\includegraphics[width=9.0 cm]{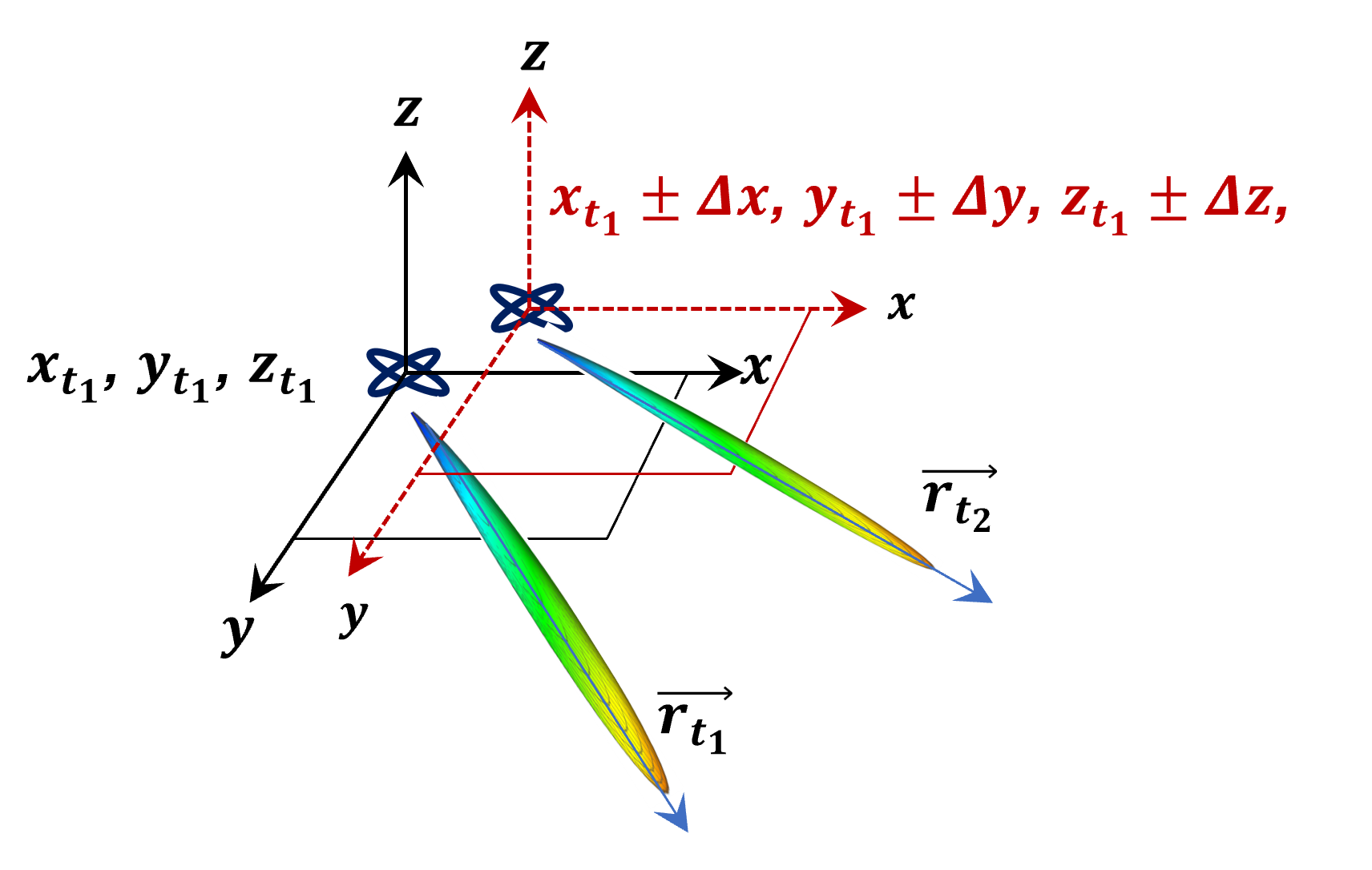}
\caption{Displacement noise due to hovering.\label{FIG-2AA}}
\end{figure}

\begin{figure}
    \centering
    \begin{subfigure}[b]{0.51\textwidth}
        \centering
        \includegraphics[height=2.4in]{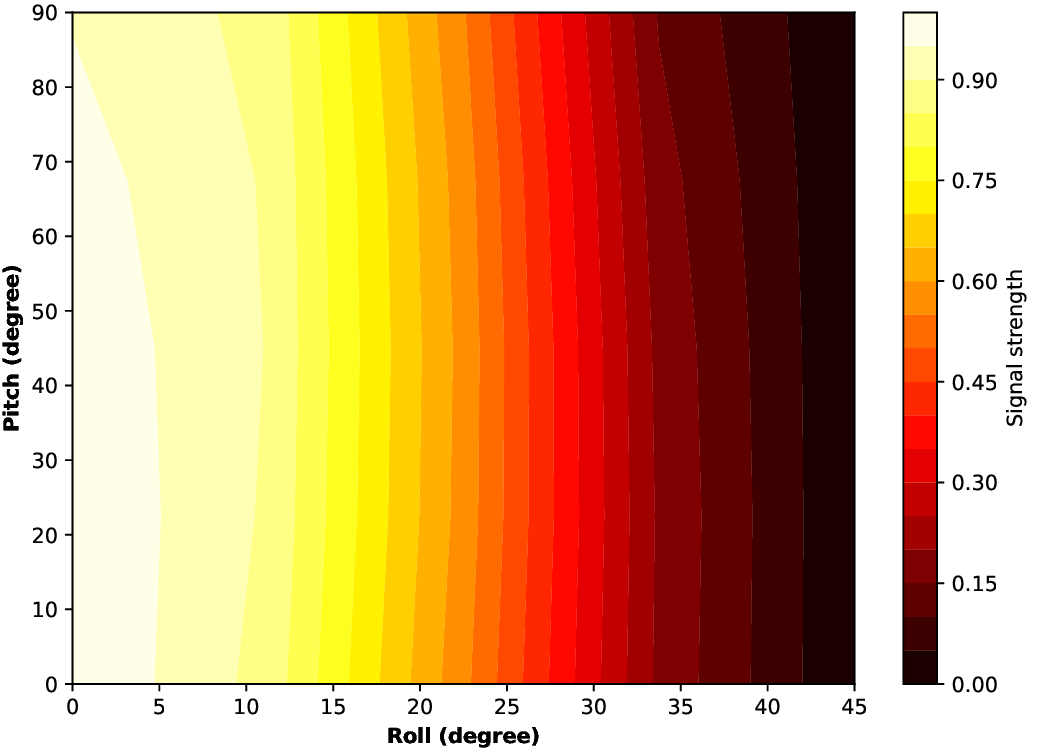}
        \caption{Beam distortion and received power variation.}
        \label{FIG-6A}
    \end{subfigure}%
    
    \begin{subfigure}[b]{0.51\textwidth}
        \centering
        \includegraphics[height=2.4in]{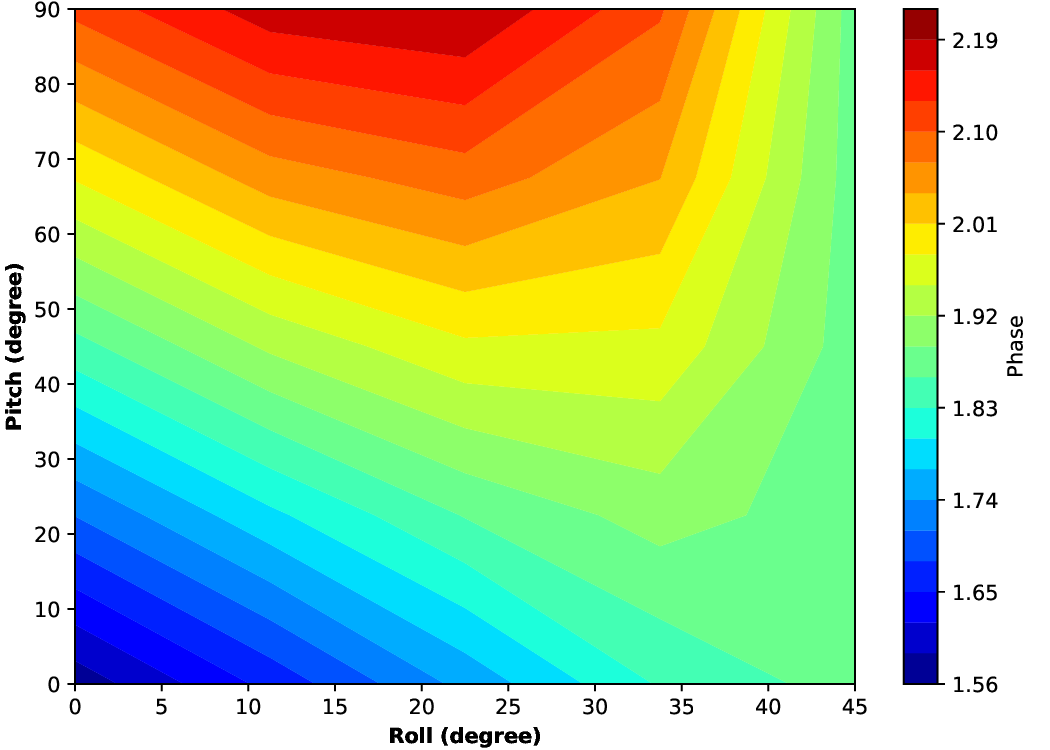}
        \caption{Phase distortion.}
        \label{FIG-6B}
    \end{subfigure}
    
    \caption{Performance impairment due to UAVs hovering, $\{R_X, R_Y, R_Z\} = \{0, 0, 50\}, \xi = 0$ degree.}
    \label{FIG-6}
\end{figure}

\begin{figure}
    \centering
    \begin{subfigure}[b]{0.51\textwidth}
        \centering
        \includegraphics[height=2.4in]{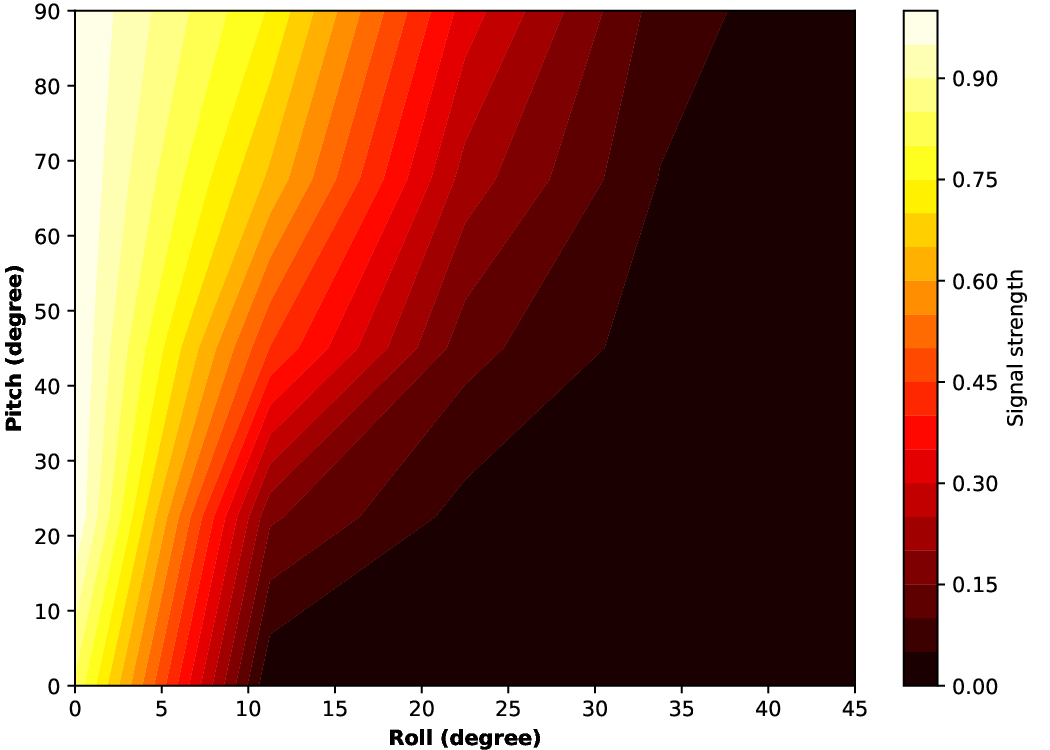}
        \caption{Beam distortion and received power variation.}
        \label{FIG-7A}
    \end{subfigure}%
    
    \begin{subfigure}[b]{0.51\textwidth}
        \centering
        \includegraphics[height=2.4in]{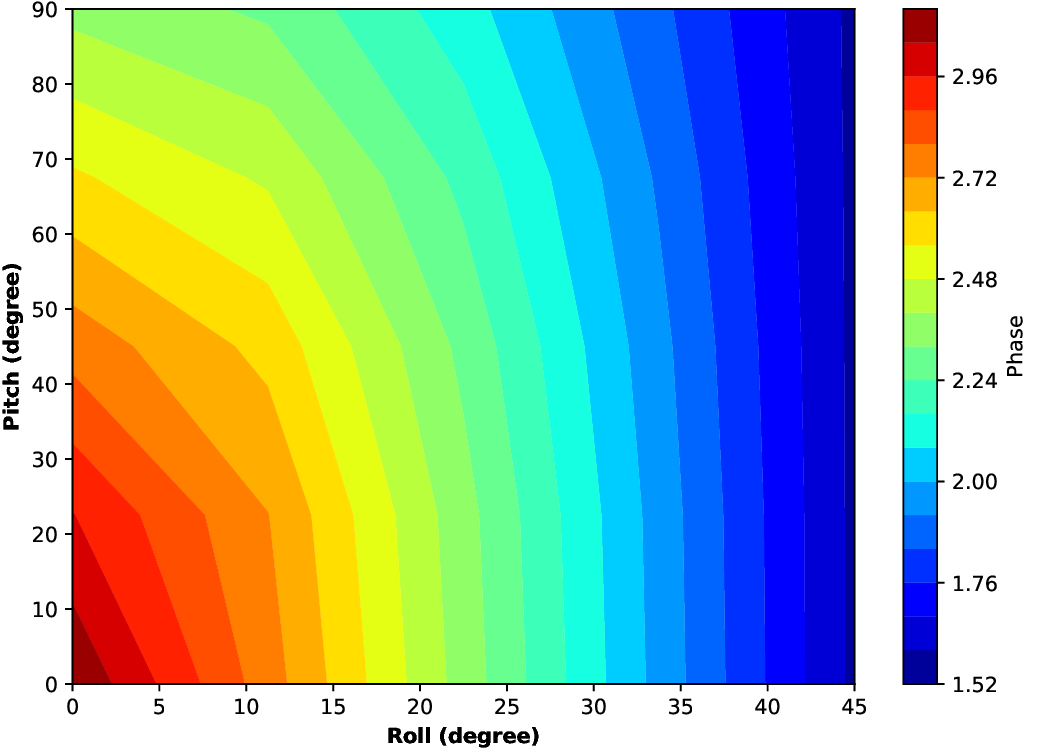}
        \caption{Phase distortion.}
        \label{FIG-7B}
    \end{subfigure}
    
    \caption{Performance impairment due to UAVs hovering, $\{R_X, R_Y, R_Z\} = \{0, 0, 100\}, \xi = 0$ degree.}
    \label{FIG-7}
\end{figure}

\begin{figure}
    \centering
    \begin{subfigure}[b]{0.51\textwidth}
        \centering
        \includegraphics[height=2.4in]{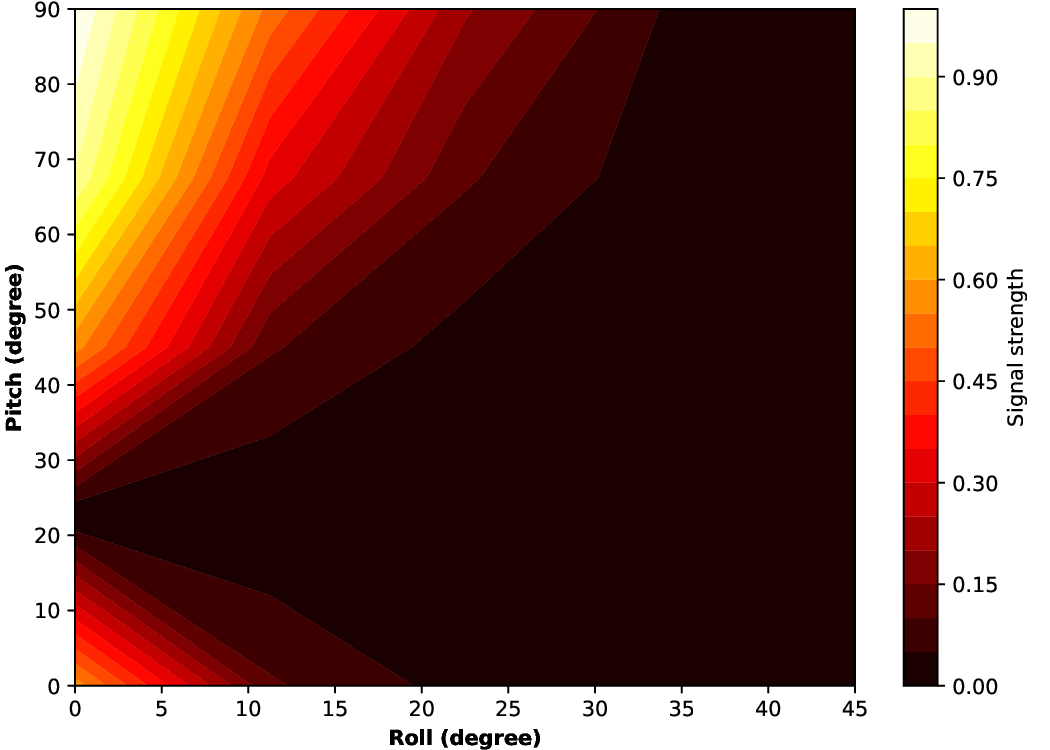}
        \caption{Beam distortion and received power variation.}
        \label{FIG-8A}
    \end{subfigure}%
    
    \begin{subfigure}[b]{0.51\textwidth}
        \centering
        \includegraphics[height=2.4in]{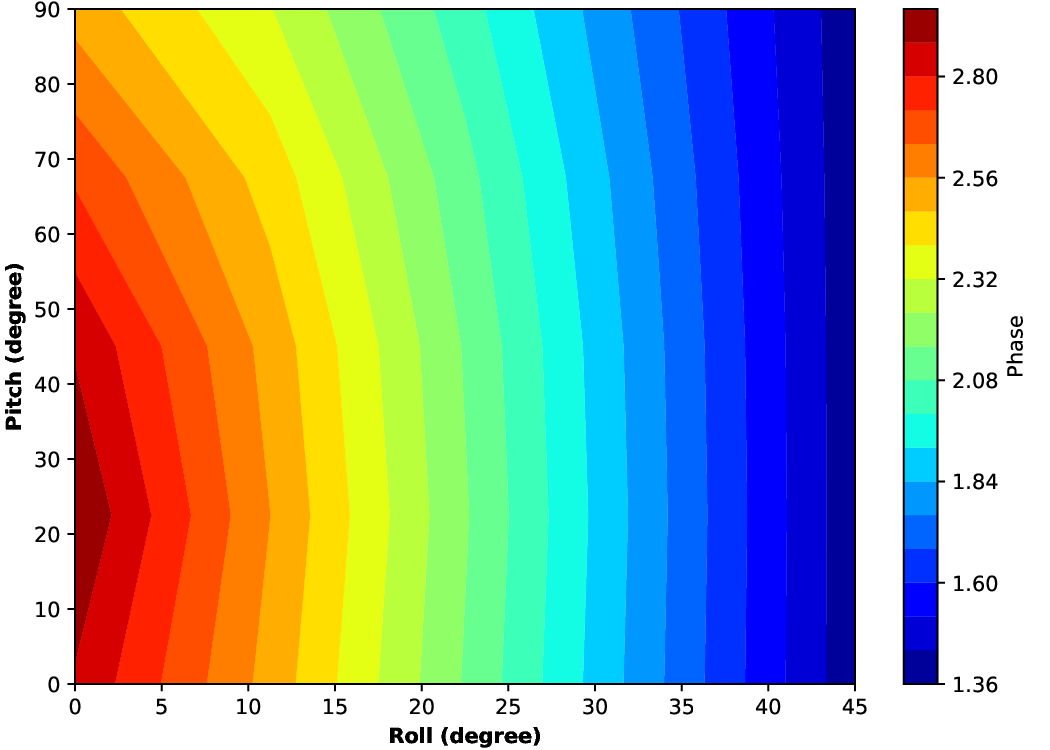}
        \caption{Phase distortion.}
        \label{FIG-8B}
    \end{subfigure}
    
    \caption{Performance impairment due to UAVs hovering, $\{R_X, R_Y, R_Z\} = \{50, 50, 300\}, \xi = 0$ degree.}
    \label{FIG-8}
\end{figure}

Next, we show the impact of the UAVs hovering on the beamforming performance. Figures \ref{FIG-6}, \ref{FIG-7}, and \ref{FIG-8} show the received power variations and phase distortions relative to the different values of the pitch angle $\gamma$ and the roll angle $\Theta$ for different receiver locations, i.e., is, $\{x_R, y_R, z_R, \} = \{0, 0, 50\}, \{0, 0, 100\}$, and $\{50, 50, 300\}$ (meters), respectively. In obtaining the results in Figs.  \ref{FIG-6}, \ref{FIG-7}, and \ref{FIG-8}, the pitch angle $\xi$ is maintained constant with value set to $0$ degree. For brevity, we consider four UAVs arranged in a linear array at locations $\{x_k, y_k, z_k \} = \{0, 0, 30\}, \{1, 0, 30\}, \{2, 0, 30\}, \{3, 0, 30\}$ (meters) for $k = \{1, 2, 3, 4$\}, respectively. The beamforming weights are obtained by utilizing the maximum ratio transmission (MRT) technique. The results show that the received power is very much sensitive to the variations in pitch angle $\gamma$ and roll angle $\Theta$. These results suggest that hovering can result in high variations in the received power with spurious sidelobe. It also reduces the receiver sensitivity and angle measurement ability.

\begin{figure}
    \centering
    \begin{subfigure}[b]{0.50\textwidth}
        \centering
        \includegraphics[height=1.5in]{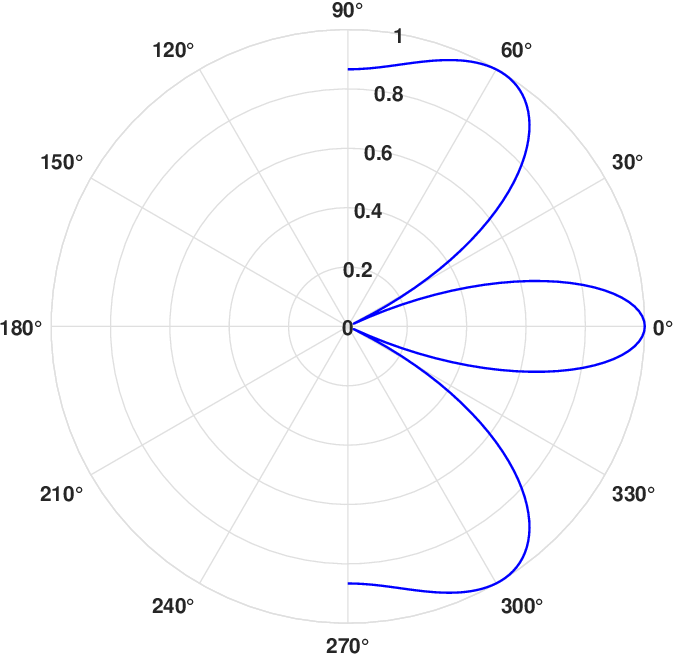}
        \caption{Beamforming pattern with $\{\Delta x, \Delta y, \Delta z\} = \pm 1$ cm.}
        \label{FIG-9A}
    \end{subfigure}%
    
    \begin{subfigure}[b]{0.50\textwidth}
        \centering
        \includegraphics[height=1.5in]{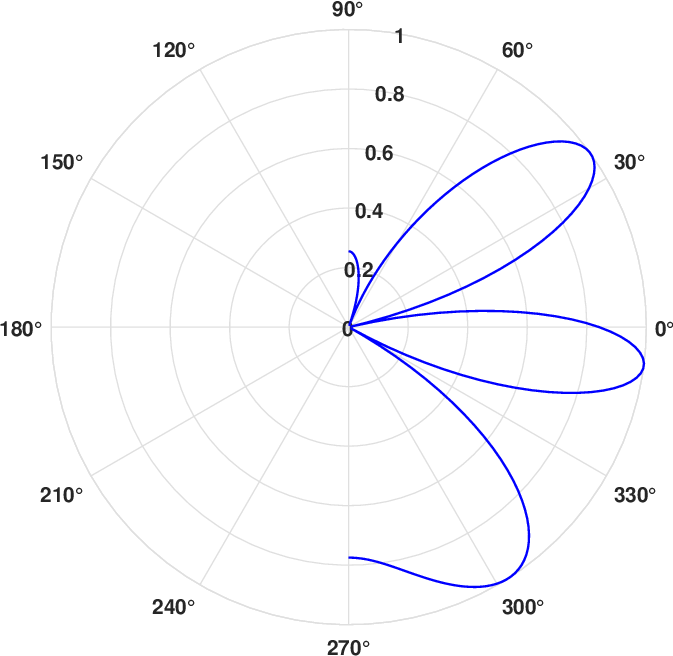}
        \caption{Beamforming pattern with $\{\Delta x, \Delta y, \Delta z\} = \pm 2$ cm.}
        \label{FIG-9B}
    \end{subfigure}

    \begin{subfigure}[b]{0.50\textwidth}
        \centering
        \includegraphics[height=1.5in]{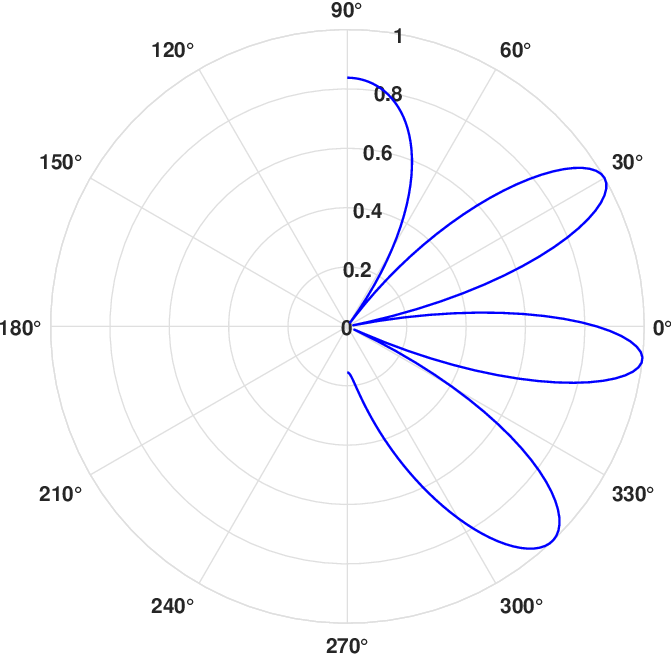}
        \caption{Beamforming pattern with $\{\Delta x, \Delta y, \Delta z\} = \pm 3$ cm.}
        \label{FIG-9C}
    \end{subfigure}

    \begin{subfigure}[b]{0.50\textwidth}
        \centering
        \includegraphics[height=1.5in]{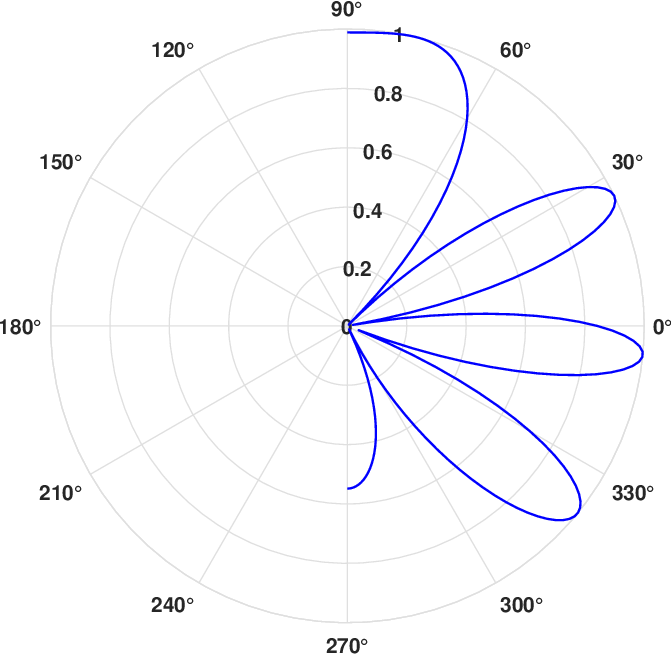}
        \caption{Beamforming pattern with $\{\Delta x, \Delta y, \Delta z\} = \pm 4$ cm.}
        \label{FIG-9D}
    \end{subfigure}

    \begin{subfigure}[b]{0.50\textwidth}
        \centering
        \includegraphics[height=1.5in]{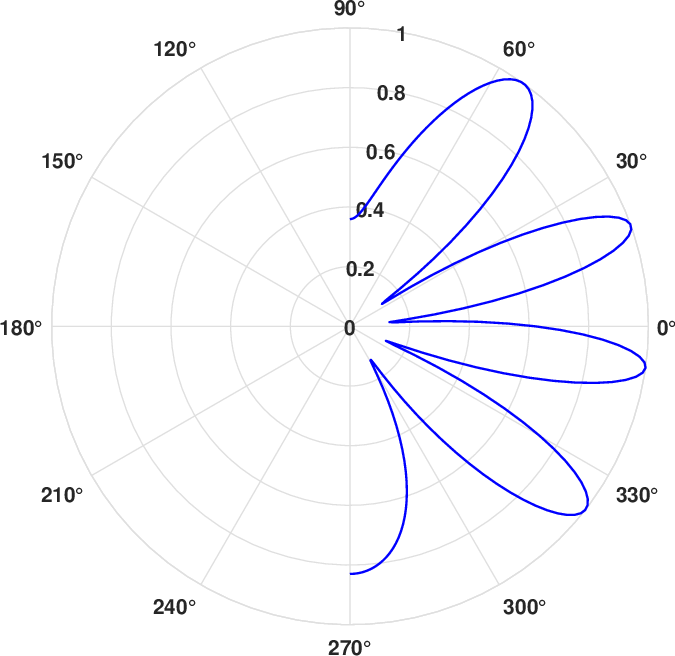}
        \caption{Beamforming pattern with $\{\Delta x, \Delta y, \Delta z\} = \pm 5$ cm.}
        \label{FIG-9E}
    \end{subfigure}
    
    \caption{Impact of the displacement noise due to UAVs hovering, $\{x_R, y_R, z_R\} = \{50, 50, 200\}$ (meters). (\textbf{Number of optimal UAVs selected:} $K = 2$).}
    \label{FIG-9}
\end{figure}

To analyze the displacement noise due to hovering and its impact on the beamforming, Fig. \ref{FIG-9} shows the beamforming distortion, with normalized gain, at the receiver end for different tolerance of $\Delta x$, $\Delta y$, and $\Delta z$. In obtaining these results, the yaw $\xi$, pitch $\gamma$, and roll $\Theta$ are set to $0$ degrees each. In obtaining these results, we set $K = 2$ with transmit UAVs initial location set to $\{x_k, y_k, z_k \} = \{0, 0, 30\} and \{1, 0, 30\}$ (meters). The receiver is assumed to be located at $\{50, 50, 300\}$ (meters). With transmit UAVs initially separated $1$ m apart, larger distortion is experienced for displacement tolerance $\{\Delta x$, $\Delta y$, $\Delta z\} = \pm 5$ cm. As can be seen, as the displacement tolerance increases, the number of side lobes increases. Moreover, the higher the UAV displacement tolerance, the higher the losses experienced, as the power of the main lobe is distributed to the side lobes.

\begin{figure}
    \centering
    \begin{subfigure}[b]{0.50\textwidth}
        \centering
        \includegraphics[height=1.5in]{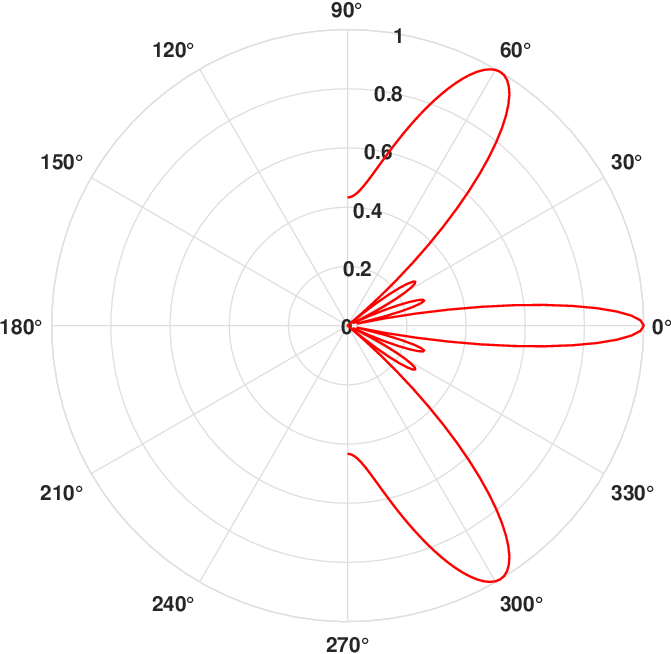}
        \caption{Beamforming pattern with $\{\Delta x, \Delta y, \Delta z\} = \pm 1$ cm.}
        \label{FIG-10A}
    \end{subfigure}%
    
    \begin{subfigure}[b]{0.50\textwidth}
        \centering
        \includegraphics[height=1.5in]{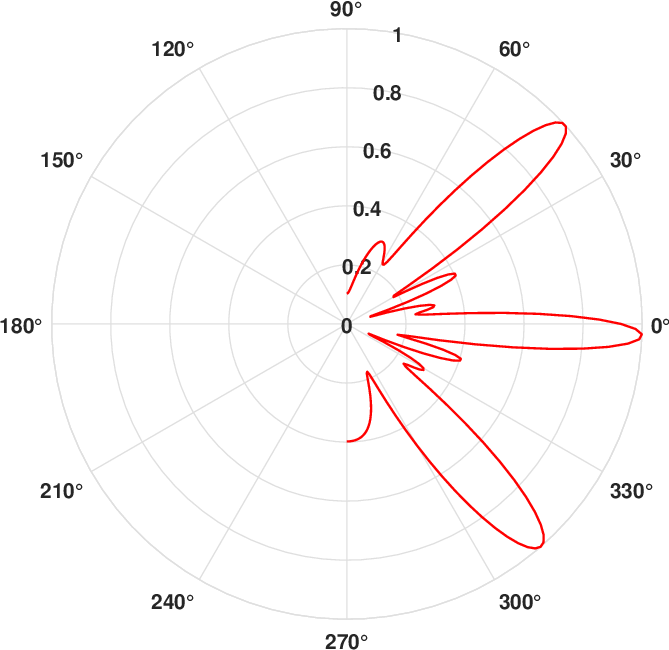}
        \caption{Beamforming pattern with $\{\Delta x, \Delta y, \Delta z\} = \pm 2$ cm.}
        \label{FIG-10B}
    \end{subfigure}

    \begin{subfigure}[b]{0.50\textwidth}
        \centering
        \includegraphics[height=1.5in]{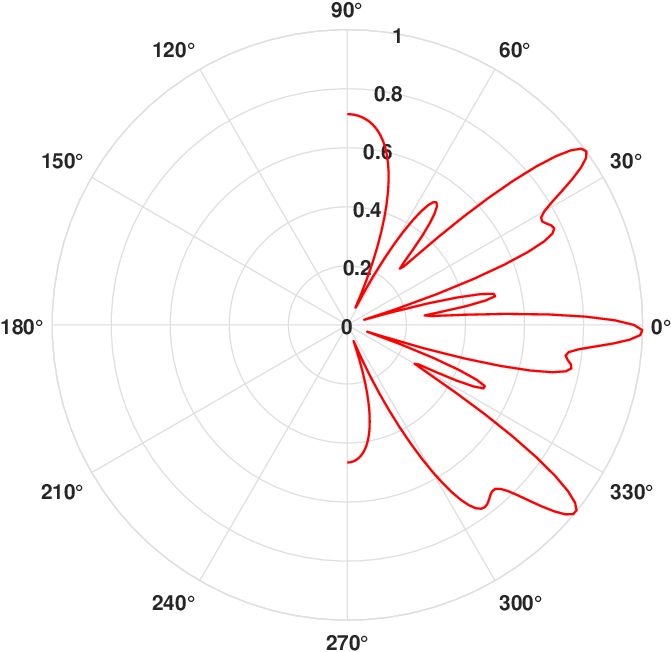}
        \caption{Beamforming pattern with $\{\Delta x, \Delta y, \Delta z\} = \pm 3$ cm.}
        \label{FIG-10C}
    \end{subfigure}

    \begin{subfigure}[b]{0.50\textwidth}
        \centering
        \includegraphics[height=1.5in]{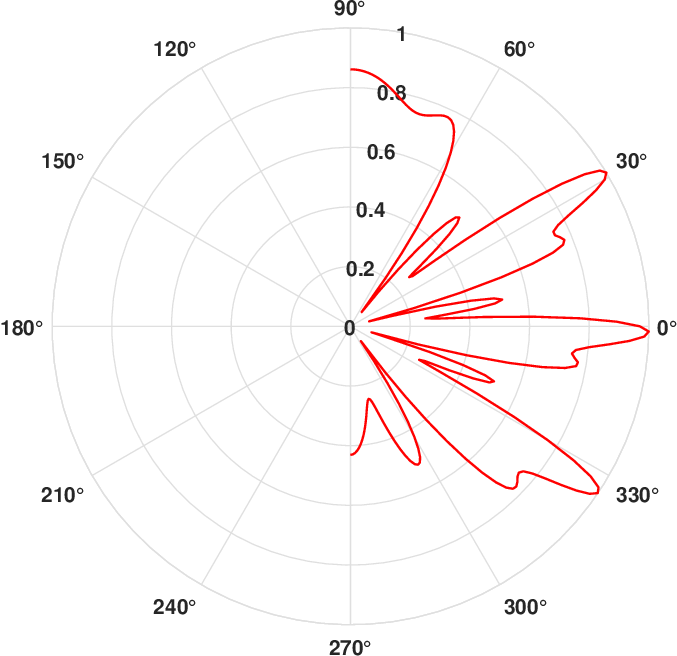}
        \caption{Beamforming pattern with $\{\Delta x, \Delta y, \Delta z\} = \pm 4$ cm.}
        \label{FIG-10D}
    \end{subfigure}

    \begin{subfigure}[b]{0.50\textwidth}
        \centering
        \includegraphics[height=1.5in]{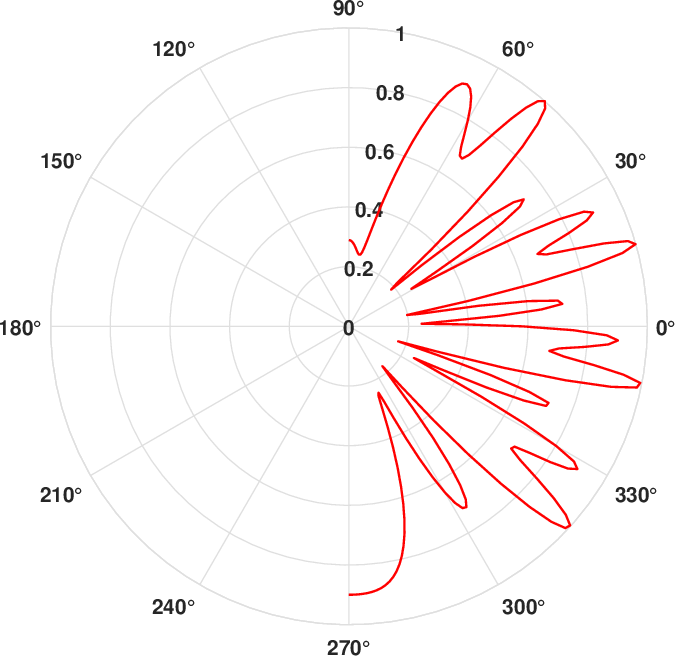}
        \caption{Beamforming pattern with $\{\Delta x, \Delta y, \Delta z\} = \pm 5$ cm.}
        \label{FIG-10E}
    \end{subfigure}
    
    \caption{Impact of the displacement noise due to UAVs hovering, $\{x_R, y_R, z_R\} = \{50, 50, 200\}$ (meters).  (\textbf{Number of optimal UAVs selected:} $K = 4$).}
    \label{FIG-10}
\end{figure}

Similar to the results obtained in Fig. \ref{FIG-9}, Fig. \ref{FIG-10} shows the impact of the displacement tolerance corresponding to $K = 4$. UAVs initial location set to $\{x_k, y_k, z_k \} = \{0, 0, 30\}, \{1, 0, 30\}, \{2, 0, 30\}, \{3, 0, 30\}$ (meters). The receiver is assumed to be located at $\{50, 50, 300\}$ (meters). It is important to note that as the number of optimal UAVs increases, the directivity of the beam increases. However, contrary to the results obtained in Fig. \ref{FIG-9}, the beam distortion also increases with displacement tolerance when $K$ increases from $2$ to $4$.

\begin{table}
	\renewcommand{\arraystretch}{1.5}
	\caption{Notations}
	\label{TABLE-I}
	\centering
	\begin{tabular}{ l p{2.5cm} } %{|c|c| } 
		\hline
		\textbf{Parameter} & \textbf{Description} \\
        \hline
        $L_U$ & Number of UAVs along $X-$axis \\
        \hline
        $C_U$ & Number of UAVs along $Y-$axis \\
		\hline
        $R_U$ & Number of UAVs along $Z-$axis \\
        \hline
        $K, K \leq L_U \times C_U \times R_U$ & Number of optimal UAVs selected for beamforming \\
        \hline
        $\mathbf{s(t)}$ & Transmitted signal \\
        \hline
		$x_k, y_k, z_k; k \in \{1,\cdot\cdot\cdot, K\}$ & 3D coordinates of the $k$th UAV\\
        \hline
		$x_R, y_R, z_R$ & 3D coordinates of the UE (receiver)\\
		\hline
            $P_k$ & Transmit power of the $k$th UAV\\
		\hline
            $\zeta_k$ & Phase of the transmitted signal of the $k$th UAV\\
		\hline
            $\theta$ & Elevation angle \\
            \hline
            $\varphi$ & Azimuthal angle \\
            \hline
            $f_c = 3.5$ GHz $(\lambda)$ & Carrier frequency (wavelength) \\
            \hline
            $\delta$    & Spacing between the adjacent UAVs \\
            \hline
            $\Delta x, \Delta y, \Delta z$    &    Displacement due to hovering along $x, y, $ and $z$ directions, respectively \\
            \hline
            $\xi, \gamma, \Theta$    &    Yaw, Pitch, and Roll (Rotational motion due to hovering) \\
            \hline
    \end{tabular} 
\end{table}

\section{Problem Formulation}
The generalized beam-pattern for $K$ UAVs located at position $\mathbf{r}_k  \buildrel \Delta \over = \left[ {x_k ,y_k ,z_k} \right]^T  \in \mathbb{R}^3, m = \{1, 2, \cdot\cdot\cdot, K\}$ and transmitting with powers $P_1, P_2, \cdot\cdot\cdot, P_K$ and phases $\zeta_1, \zeta_2, \cdot\cdot\cdot, \zeta_K$ can be expressed as
\begin{equation}
    \begin{split}
& B_{\theta ,\varphi } \left( {\mathbf{r},\mathbf{P},\mathbf{\zeta} } \right) = \left| {\sum\limits_{k = 1}^K {P_k w_k \exp \left[ {j\left( {\zeta _k  + \frac{{2\pi }}{\lambda }x_k \cos \varphi \sin \theta } \right.} \right.} } \right. \\ 
& \hspace{1.8cm}\left. {\left. {\left. { + \frac{{2\pi }}{\lambda }y_k \sin \varphi \sin \theta  + \frac{{2\pi }}{\lambda }z_k \cos \theta } \right)} \right]} \right| \\
    \end{split}
    \label{EQ-1},
\end{equation}
where $\lambda$ is the carrier wavelength, $\mathbf{r} = \left[\mathbf{r}_1^T, \mathbf{r}_2^T, \cdot\cdot\cdot, \mathbf{r}_K^T\right] \in \mathbb{R}^{3K}$. $\theta \in [-\pi, \pi]$ and $\varphi \in [-\pi, \pi]$ represent the elevation and the azimuthal angles, respectively.

We characterize the beam distortion and misalignment as system-induced errors, which include positioning and synchronization errors due to the hovering and rotational motion, and channel-induced errors, including phase distortion and interference. These errors can be modeled as perturbations in the location and the orientation parameters $\{\Delta x, \Delta y, \Delta z, \Delta \theta, \Delta \varphi\}$, and the phase parameter $\Delta\zeta$. The distorted beam pattern can then be written as
\begin{equation}
    \begin{split}
& \hat B_{\hat \theta ,\hat \varphi } \left( { \mathbf{\hat r},\mathbf{P}, \mathbf{\hat \zeta }} \right) = \left| {\sum\limits_{k = 1}^K {P_k w_k \exp \left[ {j\left( {\hat \zeta _k  + \frac{{2\pi }}{\lambda }\hat x_k \cos \hat \varphi \sin \hat \theta } \right.} \right.} } \right. \\ 
& \hspace{1.8cm} \left. {\left. {\left. { + \frac{{2\pi }}{\lambda }\hat y_k \sin \hat \varphi \sin \hat \theta  + \frac{{2\pi }}{\lambda }\hat z_k \cos \hat \theta } \right)} \right]} \right| \\ 
    \end{split}
    \label{EQ-2},
\end{equation}
where ${\hat \zeta } = \zeta \pm \Delta\zeta$, ${\hat{x}_k} = x_k \pm \Delta x_k$, ${\hat{y}_k} = y_k \pm \Delta y_k$, ${\hat{z}_k} = z_k \pm \Delta z_k$, ${\hat{z}_k} = z_k \pm \Delta z_k$, ${\hat{\theta}_k} = \theta_k \pm \Delta \theta_k$, and ${\hat{\varphi}_k} = \varphi_k \pm \Delta \varphi_k$.

We construct an optimization problem where distributed beamforming where each UAV adjusts its coordinates $x_k, y_k, z_k$, rotational angles $\xi, \gamma, \Theta$, and phase $\zeta_k$ for $k \in \{1, 2, \cdot\cdot\cdot, K\}$. With this aim, the optimization problem can be written as the minimization of the following objective function:
\begin{equation}
    \begin{split}
 & \mathop {\min }\limits_{\mathbf{r},\mathbf{\Xi} ,\zeta } J\left( {\mathbf{r},\mathbf{\Xi},\zeta } \right) =  \\ 
 & \frac{1}{4}\int\limits_{ - \pi }^\pi  {\int\limits_{ - \pi }^\pi  {\left\| {B_{\theta ,\varphi } \left( {\mathbf{r},\mathbf{P},\mathbf{\zeta} } \right) - \hat B_{\hat \theta ,\hat \varphi } \left( { \mathbf{\hat r},\mathbf{P}, \mathbf{\hat \zeta }} \right)} \right\|_2^2 d\theta d\varphi } }  \\
    \end{split}
    \label{EQ-3}.
\end{equation}

\begin{figure*}
    \centering
\includegraphics[width=13.5 cm]{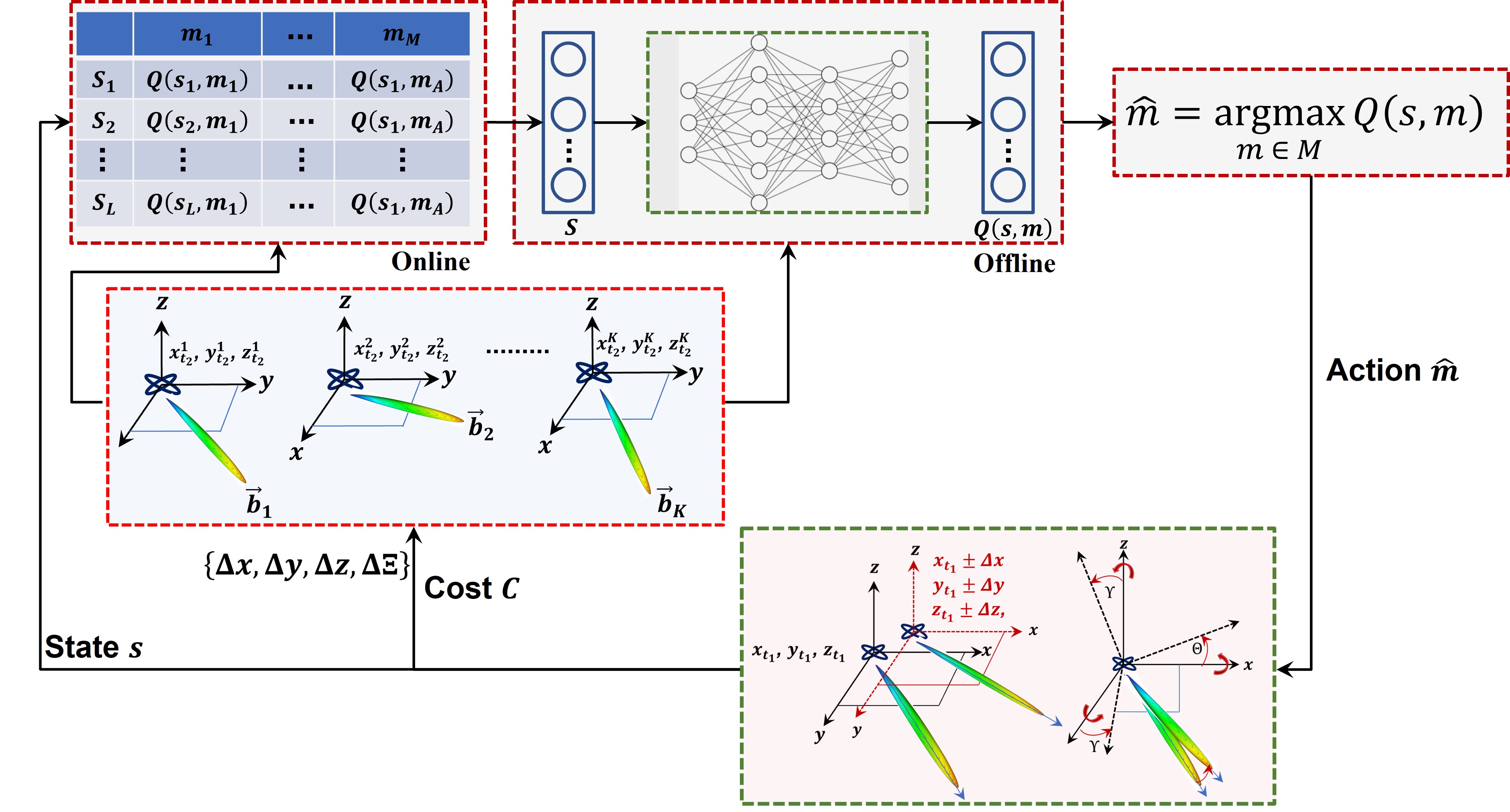}
\caption{Real-time beam tracking.\label{FIG-AA}}
\hrulefill
\end{figure*}

\section{Proposed Learning Algorithm}

\begin{algorithm}
	%\textsl{}\setstretch{1.8}
	\renewcommand{\algorithmicrequire}{\textbf{Input:}}
	\renewcommand{\algorithmicensure}{\textbf{Output:}}
	\caption{Stage I: Brute Force for Optimal Selection}
        \textbf{Input}: Total number of UAVs: $N$ \\
        \textbf{Input}: Desired number of UAVs for beamforming: $K$ \\
        \textbf{Input}: CQI at time $t$: $\mathbf{H}_t = [h_t^1, h_t^2, \cdot\cdot\cdot, h_t^N]^T$ \\
        \textbf{Define}: Empty list: Optimal\_Combination $C_{opt}^K = [\hspace{0.1cm}]$ \\
        \textbf{Initialize}: Optimal performance:  $F_{Max} \leftarrow - \infty$ \\
        \begin{algorithmic}[1] 
        \STATE UAV\_Index = list(range(1, $N+1$)) \\
        \STATE Generate an iterator that produces tuples of length $K$ representing all possible combinations of $K$ UAVs, i.e., $\{c_1^K, c_2^K, \cdot\cdot\cdot, c_{\mathcal{M}}^K\} \in \mathcal{C}$, from the list UAV\_Index: \\
        $\mathcal{C} = $itertools.combinations(UAV\_Index, $K$)
        \FOR{$j=1$ to $\mathcal{M}$}
            \STATE Obtain the CQI submatrix $\mathbf{H}_{t,c_j^K} \in \mathbf{H}_t$ corresponding to the combination $c_j^K$
            \STATE Obtain the weight vector:\\
            $ \mathbf{w}_{t, c_j^K} = \frac{\mathbf{H}_{t,c_j^K}^H}{||\mathbf{H}_{t,c_j^K}^H||}$ \\
            \STATE Calculate SINR: $F$
            \IF{$F > F_{Max}$}
                \STATE $C_{opt}^K = c_j^K$
                \STATE $F_{Max} = F$
            \ENDIF
        \ENDFOR
        \RETURN $C_{opt}^K$
        \end{algorithmic}
        \label{ALGO-1}
\end{algorithm}

\begin{figure}
    \centering
\includegraphics[width=8.5 cm]{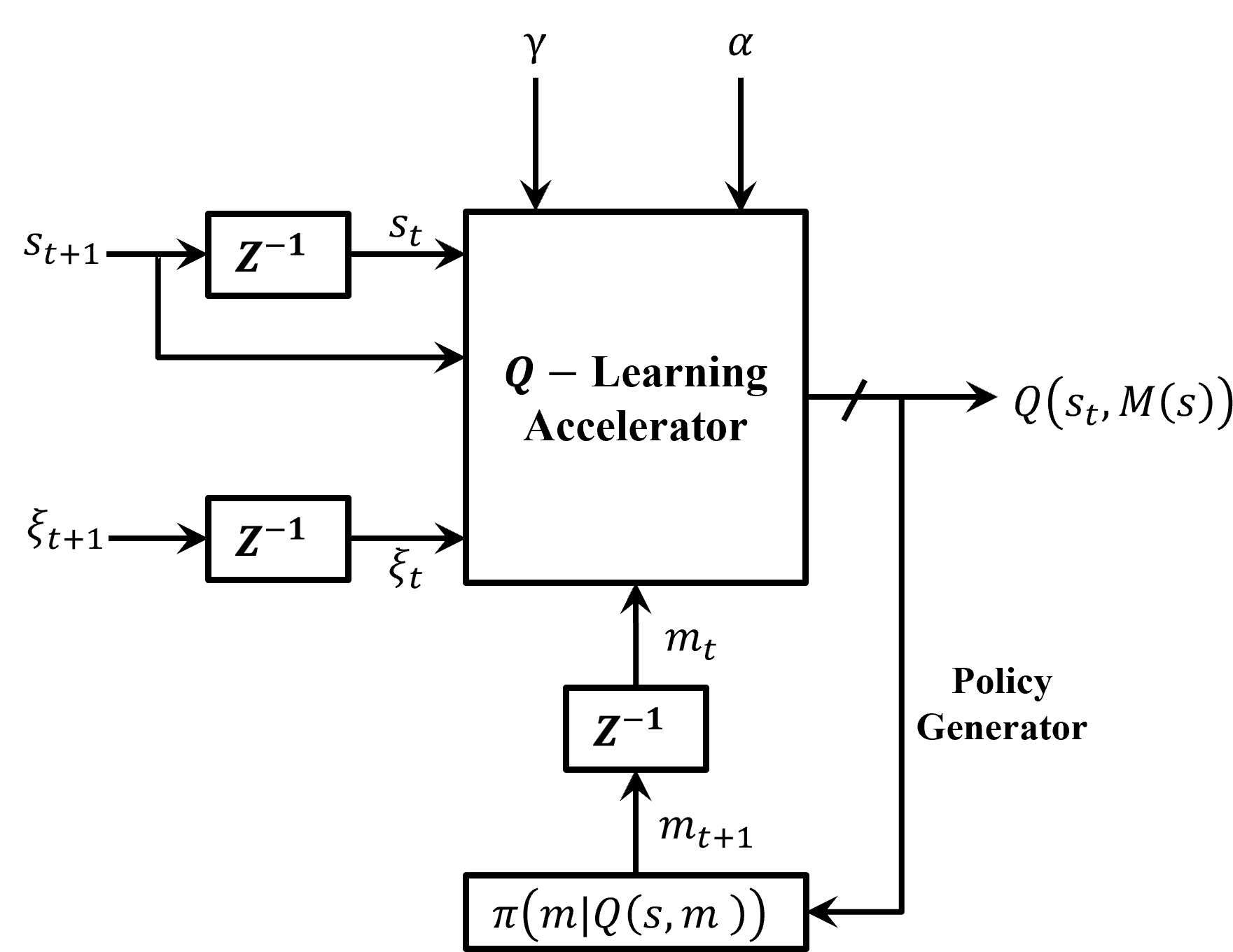}
\caption{Structural design of the $Q$-Learning agent.\label{FIG-ZZ}}
\end{figure}

We consider a deep neural network to estimate the $Q-$value corresponding to each action in each state. These estimates are then updated using the Bellman equation. An epsilon-greedy policy is utilized to balance exploration and exploitation during the training phase. A mean squared error (MSE) loss function is considered to train the neural network by measuring how well the predicted $Q-$values match the target $Q-$values. The target $Q-$values are obtained using the Bellman equation as \cite{sutton2018reinforcement}
\begin{equation}
    \begin{split}
& Q\left( {s,m} \right) =  \\ 
& Q\left( {s,m} \right) + \alpha \left[ {\xi  + \gamma \mathop {\arg \max }\limits_{m' \in \mathcal{M}\left( s \right)} \left[ {Q\left( {s',m'} \right) - Q\left( {s,m} \right)} \right]} \right] \\
    \end{split},
\end{equation}
where $\gamma \in (0, 1]$ represents a discount factor. $\gamma$ signifies the importance of long-term rewards compared with the present rewards. $\xi$ is the current reward.
\begin{remark}
It is to be noted that the concept of a terminal state and setting $Q-$values to $0$ in the terminal state does not apply to the environment presented in Algorithm \ref{ALGO-1}. We consider a problem where the episode termination is based on a specific condition, that is, $\eta \leq \eta_{Th}$.
\end{remark}
The detailed structural design of the $Q$-Learning agent is illustrated in Fig. \ref{FIG-ZZ}.

\begin{table}
	\renewcommand{\arraystretch}{1.4}
	\caption{Notations}
	\label{TABLE-I}
	\centering
	\begin{tabular}{ l p{2.5cm} } %{|c|c| } 
		\hline
		\textbf{Parameter} & \textbf{Description} \\
		\hline
            ${L}$ & Number of states \\
            \hline
            $\gamma$ & Discount factor \\
            \hline
            $\xi_t$ & Reward at time $t$ \\
            \hline
            $\alpha = 0.05$ & Learning rate \\
            \hline
\end{tabular} 
\end{table}

\begin{algorithm}
	%\textsl{}\setstretch{1.8}
	\renewcommand{\algorithmicrequire}{\textbf{Input:}}
	\renewcommand{\algorithmicensure}{\textbf{Output:}}
	\caption{Stage II: Deep Q-Learning for Beam Reforming}
        \textbf{Input}: Learning rate: $\alpha \in (0, 1]$ \\
        \textbf{Input}: Policy: $\pi(m | Q(s, m))$ \\
        \textbf{Initialize}: $Q(s,m) \leftarrow 0, \forall s \in \mathcal{S}, m \in \mathcal{M}(s)$ \\
        \begin{algorithmic}[1] 
        \STATE Evaluate the objective function as: \\
        $ \Bar{R}_{\rm ini} = \min_{R} \frac{\eta}{2} {\lVert \Bar{\mathbf{\Phi}}-(R_{\rm ini}+\zeta)} \lVert_2^2$ \\
        \STATE Evaluate: \\
        $\Bar{R}_{\max} = \max{(\Bar{R}_{\rm ini})}$ \\
        $\Bar{R}_{\min}= \min(\Bar{R}_{\rm ini})$ \\
        \FOR{$j=2$ to $N-1$}
            \STATE $\Bar{R}_{\rm Th}  = \Bar{R}_{\rm Th} + \Bar{R}(j)$
            \STATE Arrange the simplex in geometric polygonal shape such that:\\
            $ \Bar{r}_{\min}< \Bar{r}(1) < \Bar{r}(2) < \cdot\cdot\cdot < \Bar{r}_{\max}$ \\
        \ENDFOR
        \WHILE{$\eta > \eta_{Th}$}
           \STATE Compute: $m \leftarrow \mathop {\arg \max }\limits_{m' \in \mathcal{M}} Q_p\left( {s,m'; \mathbf{w}} \right)$
           \STATE Compute: The target $Q-$value: \\
           $Q\left( {s,m} \right)$ = \\
            $ Q\left( {s,m} \right) + \alpha \left[ {\xi  + \gamma \mathop {\arg \max }\limits_{m' \in \mathcal{M}\left( s \right)} \left[ {Q\left( {s',m'} \right) - Q\left( {s,m} \right)} \right]} \right] $
            \STATE Compute: DQN Loss Function: \\
            $\mathcal{L}\left( \mathbf{w} \right) = \mathbb{E}_\pi  \left[ {\left( {Q\left( {s',m'} \right) - Q\left( {s',m';w} \right)} \right)^2 } \right]$
        \ENDWHILE
         \STATE \textbf{Terminate if condition met}
         \RETURN $R$
        \end{algorithmic}
        \label{ALGO-2}
\end{algorithm}

\begin{figure}
    \centering
\includegraphics[width=8.5 cm]{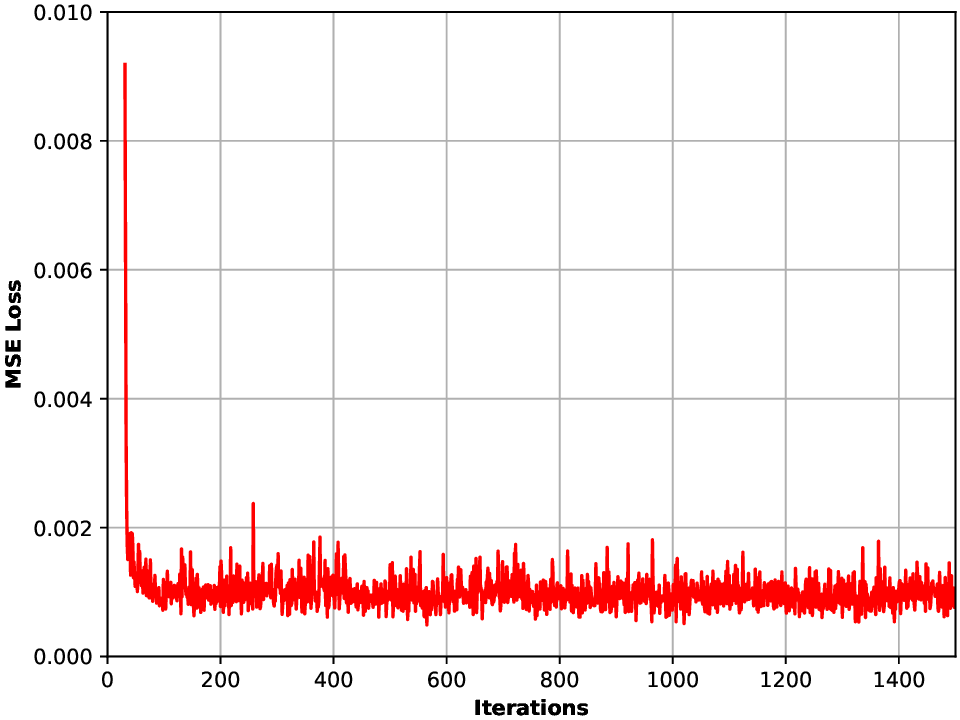}
\caption{Convergence of the MSE loss with iterations.\label{FIG-13}}
\end{figure}

To train the neural network, we consider the DQN loss function as the squared difference between the actual and the predicted action values. This is typically implemented as the mean-squared error (MSE) function. The convergence of the MSE loss function against the number of iterations is illustrated in Fig. \ref{FIG-13}. As can be readily seen, converges very quickly and learns to approximate the target values efficiently. It can be seen that with $N = 64$ and $K$ set to $4$, the proposed approach requires approximately $50$ iterations only for convergence. A sharp steeper initial slope suggests that the proposed learning algorithm fine-tunes its parameters quickly without observing any plateaus. 

To gain more insight into the proposed learning algorithm, Fig. \ref{FIG-14} shows the average reward received relative to the number of episodes. These curves are obtained for $N = 64$ and $K = 4$ with $\delta$ set to $1$ m. It can be seen that the stable average reward after $200$ episodes suggests that the network reaches its learning limit irrespective of its displacement tolerance values $\Delta$.

\begin{figure}
    \centering
\includegraphics[width=8.5 cm]{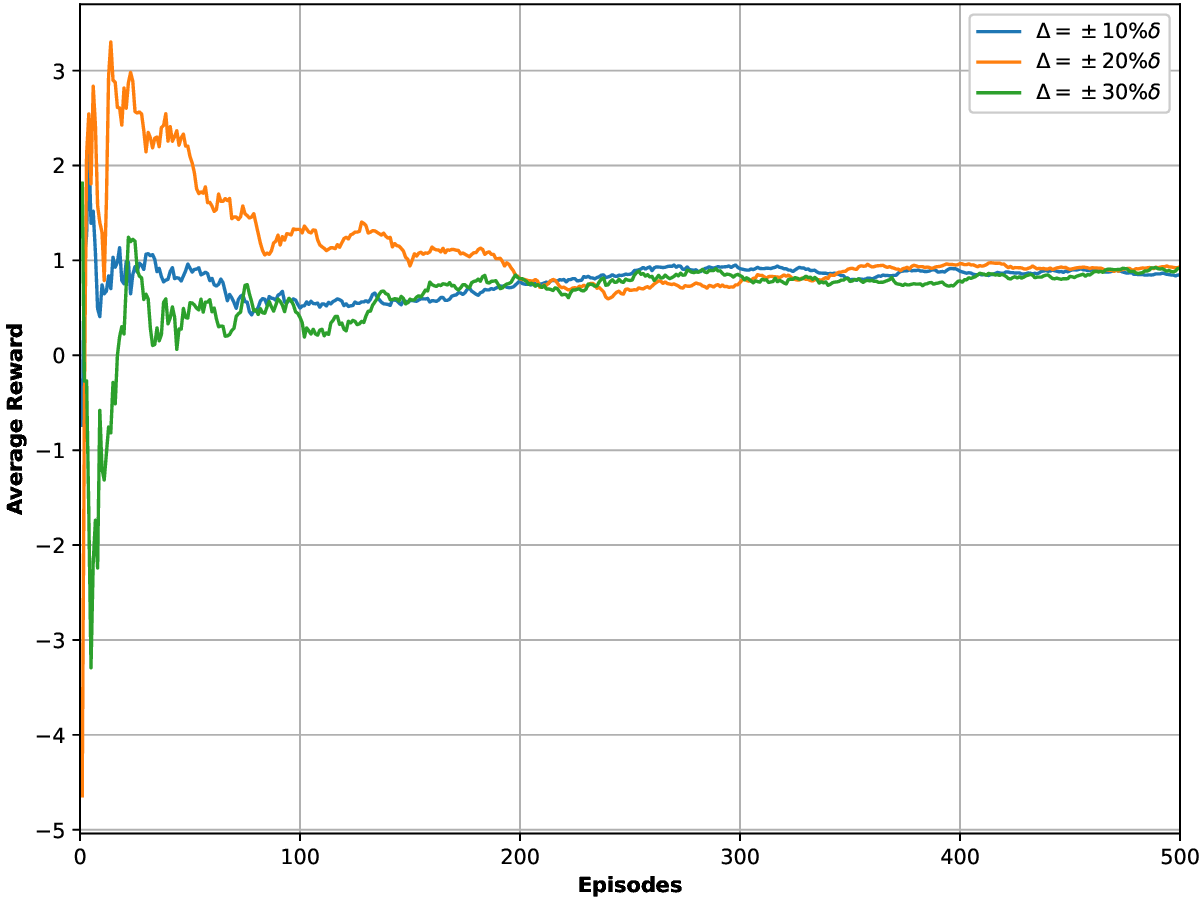}
\caption{Average reward against different hovering tolerance.\label{FIG-14}}
\end{figure}

\begin{remark}
    It is important to note that there is no explicit feedback or recurrent connections provided in the neural network architecture illustrated in {\rm Fig. \ref{FIG-AA}}. The DNN considered in the architecture is a feedforward neural network where the information flows in a single direction, from the input layer through hidden layers to the output layer. However, in the RL algorithm implemented, as depicted in {\rm Algorithm \ref{ALGO-2}}, there is an implicit form of feedback through the use of the Bellman equation to update the $Q$ values. The $Q$-values are updated based on the observed rewards and the estimated future rewards, creating a feedback loop that allows the agent to learn and improve its actions over time. But this feedback is incorporated within the $Q$-learning algorithm rather than being explicitly modeled as recurrent connections within the neural network architecture.
\end{remark}

To enable the DQN agent to learn from past experiences, we consider the idea of experience replay. At any time, $T$, the DQN agent has access to the history of states encountered, actions taken, transitions made, and rewards received. Mathematically, it can be represented as
\begin{equation}
\mathcal{H}_T  = \left\{ {\left( {s_t ,m_t ,\xi \left( {s_t ,m_t } \right),m_{t + 1} } \right)} \right\};T - H + 1 \le t \le T
\end{equation}
At any time $T$, the DQN weights are updated by sampling the mini-batch of size ${\mathcal{H}}$ from the experience replay. The average loss gradient can then readily be obtained as
\begin{equation}
    \begin{split}
& \overline {\nabla L\left( w \right)}  = \\
& \frac{1}{{\hat H}}\sum\limits_{i = 1}^{\hat H} {\nabla L\left( {w_T ,s_{t\left( {T,i} \right)} ,m_{t\left( {T,i} \right)} } \right)} ; T - H + 1 \le t\left( {T,i} \right) \le T \\ 
    \end{split}
    \label{EQ-A3}
\end{equation}
Utilizing (\ref{EQ-A3}), the weights of the DQN are then updated as
\begin{equation}
w_{t' + 1}  = w_{t'}  + \varpi \left( {t'} \right)\left[ {\frac{1}{{\hat H}}\sum\limits_{i = 1}^{\hat H} {\nabla L\left( {w_t ,s_{t\left( {t',i} \right)} ,m_{t\left( {t',i} \right)} } \right)} } \right]
\label{EQ-A4},
\end{equation}
where $\varpi \left( {t'} \right)$ is the step-size sequence. It is to be noted that $\varpi \left( {t'} \right)$ satisfies the standard assumptions of square summability and nonsummability.

\section{Performance Analysis}

\subsection{Impact of the Jitter and Displacement Noise due to Hoevering on the Receiver Sensitivity}
\begin{figure}
    \centering
    \begin{subfigure}[b]{0.45\textwidth}
        \centering
        \includegraphics[height=2.1in]{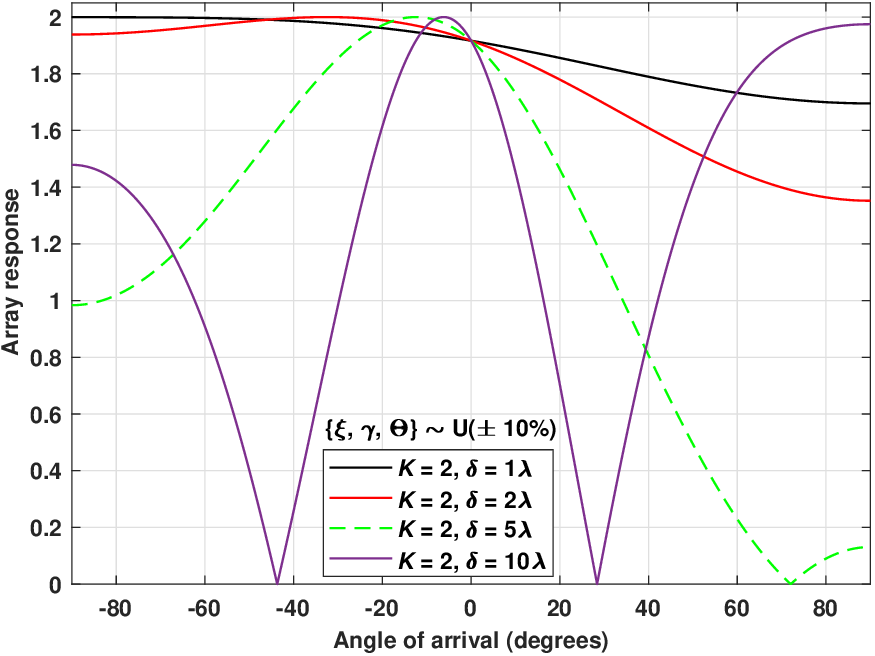}
        \caption{.}
        \label{FIG-16A}
    \end{subfigure}%
    
    \begin{subfigure}[b]{0.45\textwidth}
        \centering
        \includegraphics[height=2.1in]{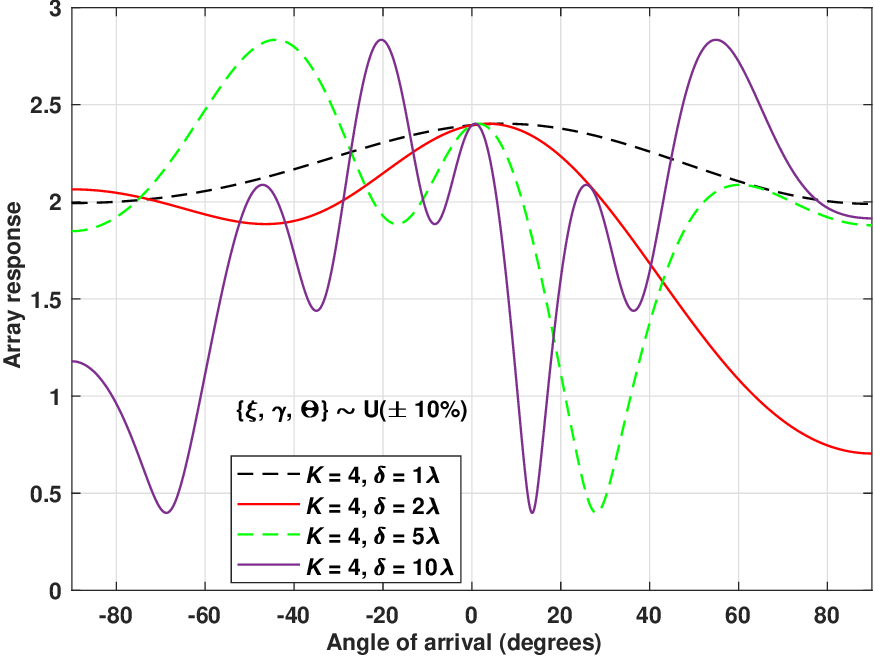}
        \caption{.}
        \label{FIG-16B}
    \end{subfigure}
    
    \caption{Array response relative to the angle of arrival for different UAVs spacing.}
    \label{FIG-16}
\end{figure}

The UAVs array response relative to the angle of arrival for different UAV spacing is illustrated in Fig. \ref{FIG-16}. The important observations that can be inferred from the results are listed below. These analyses are obtained considering the Yaw, Pitch, and Roll to be uniformly distributed within the range of $\pm 10$ degrees, whereas.
\begin{itemize}
    \item As the spacing between the UAVs in the network increases, the beamwidth of the array becomes narrow. It improves the network's ability to focus more energy in a specific direction. Note that in practice, increased directivity leads to effectively distinguishing the signals arriving from different directions at the receiver, thereby improving the receiver's ability to suppress the interference and enhance the desired signal.
    \item However, as can be seen from the results, the sidelobes increase with the spacing.
\end{itemize}

\subsection{Performance Analysis Considering Impairments due to the Hoevering and Interference from the Neighboring Networks}
The implementation of Algorithm \ref{ALGO-1} is illustrated in Fig. \ref{FIG-14}. Based on the results obtained, we observe that the performance varied across different time instants, indicating the dynamic nature of the wireless channel. In obtaining the results shown in Fig. \ref{FIG-14}, we consider a 3D rectangular geometry of the UAVs network, as illustrated in Fig. \ref{FIG-1}. As a benchmark, a brute-force search is applied over $\frac{N!}{K! \times (N - K)!}$ combinations to select $K$ best UAVs from $N$ number of UAVs arranged in a rectangular 3D geometry, to maximize the received signal-to-interference-plus-noise ratio (SINR) and minimize the latency. To generate random interference regions, we consider interference signals from randomly distributed neighboring networks. It is assumed that the neighboring interference sources are distributed uniformly in the space. While accounting for the interference from the satellites, this distribution reflects a practical scenario \cite{zhang2022effects}. With $N$ set to $64$ and $K = 4$, the result shows the importance of the proposed exhaustive search algorithm.
\begin{figure}
    \centering
\includegraphics[width=8.5 cm]{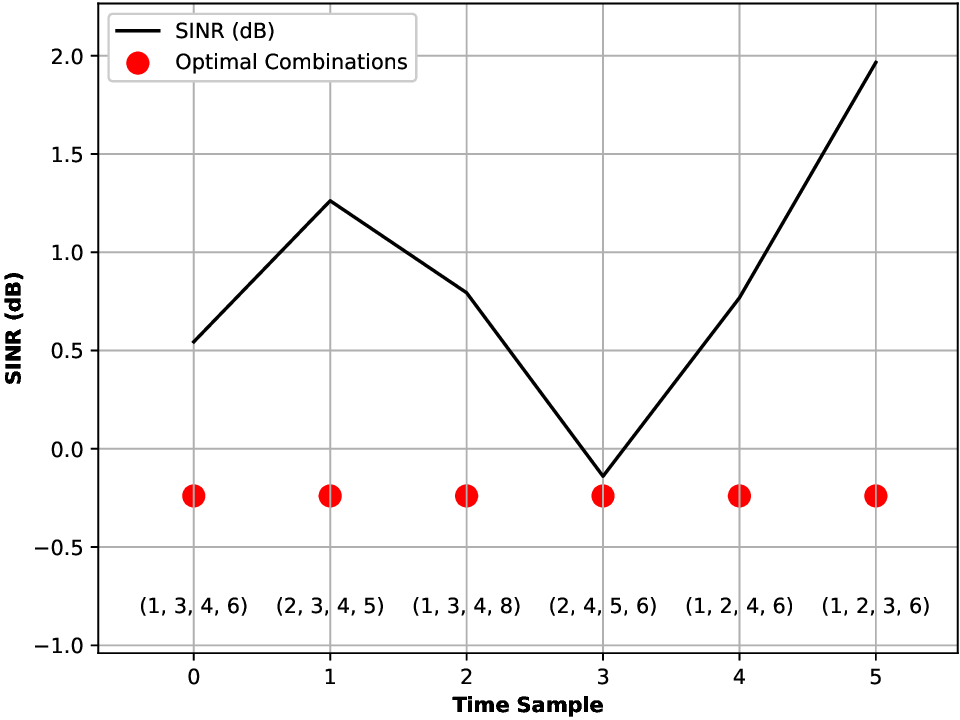}
\caption{Stage I: Illustration of brute force for Optimal UAVs selection.\label{FIG-14}}
\end{figure}

\begin{figure}
    \centering
    \begin{subfigure}[b]{0.51\textwidth}
        \centering
        \includegraphics[height=2.4in]{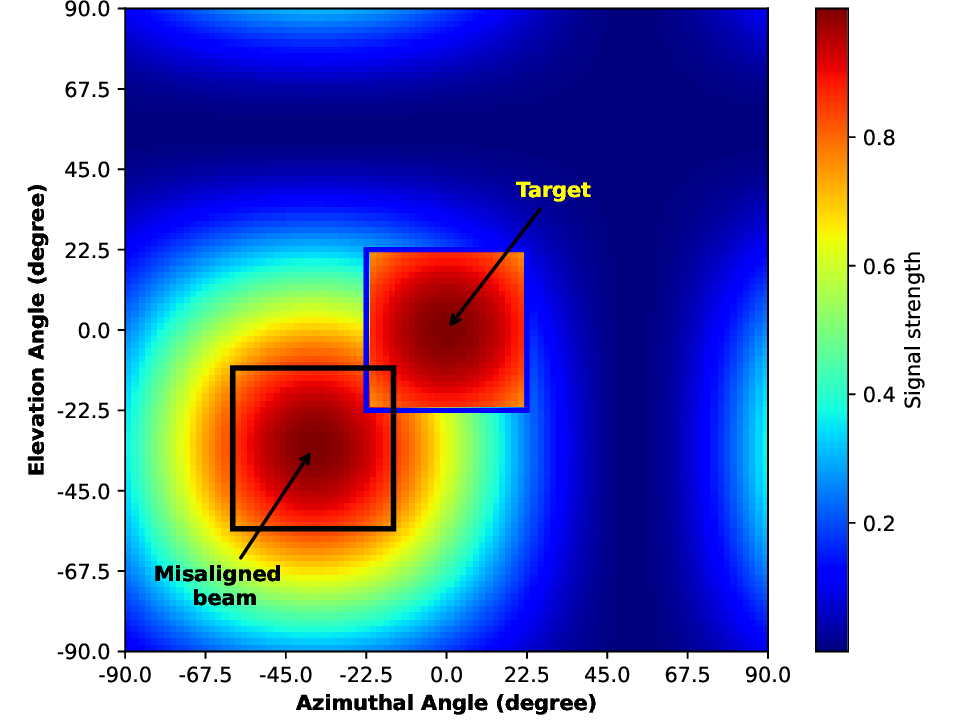}
        \caption{Beam coverage and misalignment without training and tracking.}
        \label{FIG-17A}
    \end{subfigure}%
    
    \begin{subfigure}[b]{0.51\textwidth}
        \centering
        \includegraphics[height=2.4in]{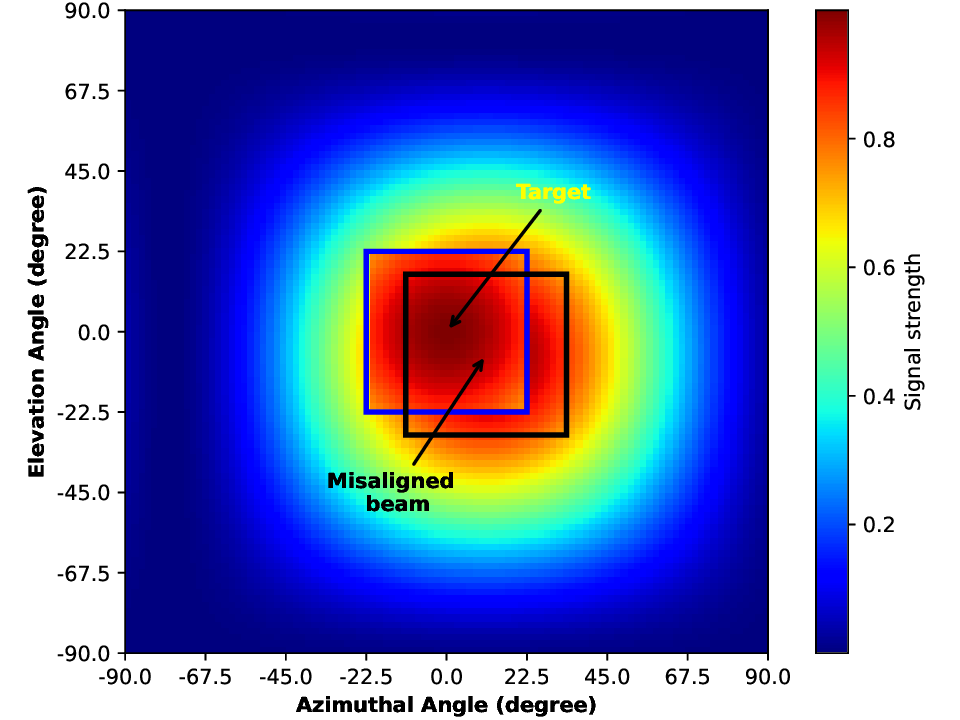}
        \caption{Beam coverage and misalignment with training and tracking.}
        \label{FIG-17B}
    \end{subfigure}
    
    \caption{Performance illustration of the proposed reinforcement learning algorithm for beam tracking.}
    \label{FIG-17}
\end{figure}

Fig. \ref{FIG-17A} depicts the misalignment or displacement of the beam center at the receiver, where the center of the directed beam points towards a slightly different location rather than the target receiver location. In obtaining these performance results, $N$ is set to $64$ whereas $K$ is set to $4$. Hovering tolerance $\Delta x, \Delta y,$ and $\Delta z$ are all set to $30\%$ of the adjacent UAVs spacing $\delta$, with $\delta$ being set to $1$ m. The 

\section{Conclusion}
Due to the inherent challenges in adaptive beam selection and beam reforming in a dynamic 3D UAV network, coexisting with the satellite and terrestrial networks, such as limitations of channel predictions due to the dynamic nature, the requirement of higher receiver sensitivity to the angle-of-arrival, and interference to/from the neighboring network, we have introduced a novel and first-of-its-kind model-free approach to the problem of dynamic and adaptive beam selection and beam reforming. To this end, first, we analyzed the impact of hovering on the beamforming performance over the interference channel. We have shown that for a UAV network, the impact of hovering increases as the number of optimal UAVs increases. For instance, it has been demonstrated that for a 3D UAV network with $64$ UAVs arranged over a rectangular geometry, with adjacent UAVs spacing set to $1$ m, the hovering tolerance of $5\%$ results in significant performance degradation when the number of optimal UAVs utilized to form a beam is set to $4$ when compared with $2$ optimal UAVs utilized to construct a beam.  We have proposed an extensive search-based optimal UAVs selection algorithm and a Deep-Q Learning based algorithm for real-time beam reforming. It has been demonstrated that the proposed collaborative beamforming approach is not only effective but also very feasible over the dynamic hovering-tolerant interference channel. Results show the fast convergence of the proposed approach. It has been shown that for a rectangular 3D geometry with $64$ UAVs, the proposed approach requires approximately $50$ iterations only for convergence when the number of UAVs required for beamforming is set to $4$. We have shown that the proposed DQN architecture with an exhaustive search algorithm fine-tunes its parameters quickly without observing any plateaus. Moreover, we have made an important observation that the learning algorithm performs efficiently and independently of the hovering tolerance values.

% if have a single appendix:
%\appendix[Proof of ]\label{APPENDIX-1}

%\appendix[Proof of ]\label{APPENDIX-2}

\ifCLASSOPTIONcaptionsoff
  \newpage
\fi

\bibliographystyle{IEEEtran}
\bibliography{IEEEabrv,Bibliography,references_ying}

\begin{IEEEbiography}[{\includegraphics[width=1.0in,height=1.25in, clip,keepaspectratio]{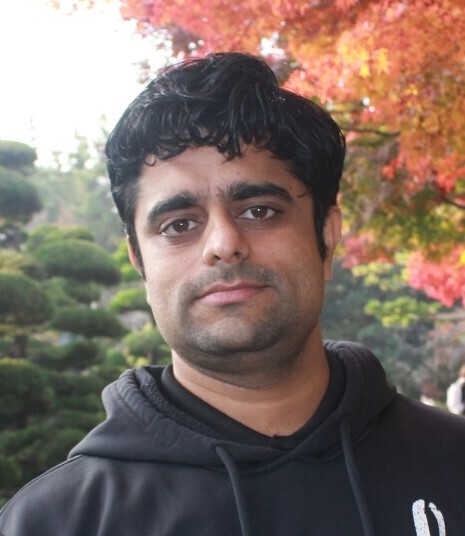}}]{Sudhanshu Arya} (Member, IEEE) is a Research Fellow in the School of System and Enterprises at Stevens Institute of Technology, NJ, USA. He received his M.Tech. degree in communications and networks from the National Institute of Technology, Rourkela, India, in 2017, and the Ph.D. degree from Pukyong National University, Busan, South Korea, in 2022. He worked as a Research Fellow with the Department of Artificial Intelligence Convergence, at Pukyong National University. His research interests include wireless communications and digital signal processing, with a focus on free-space optical communications, optical scattering communications, optical spectrum sensing, computational game theory, and artificial intelligence. He received the Best Paper Award in ICGHIT 2018 and the Early Career Researcher Award from the Pukyong National University in 2020.
\end{IEEEbiography}

\begin{IEEEbiography}[{\includegraphics[width=1in,height=1.25in, clip,keepaspectratio]{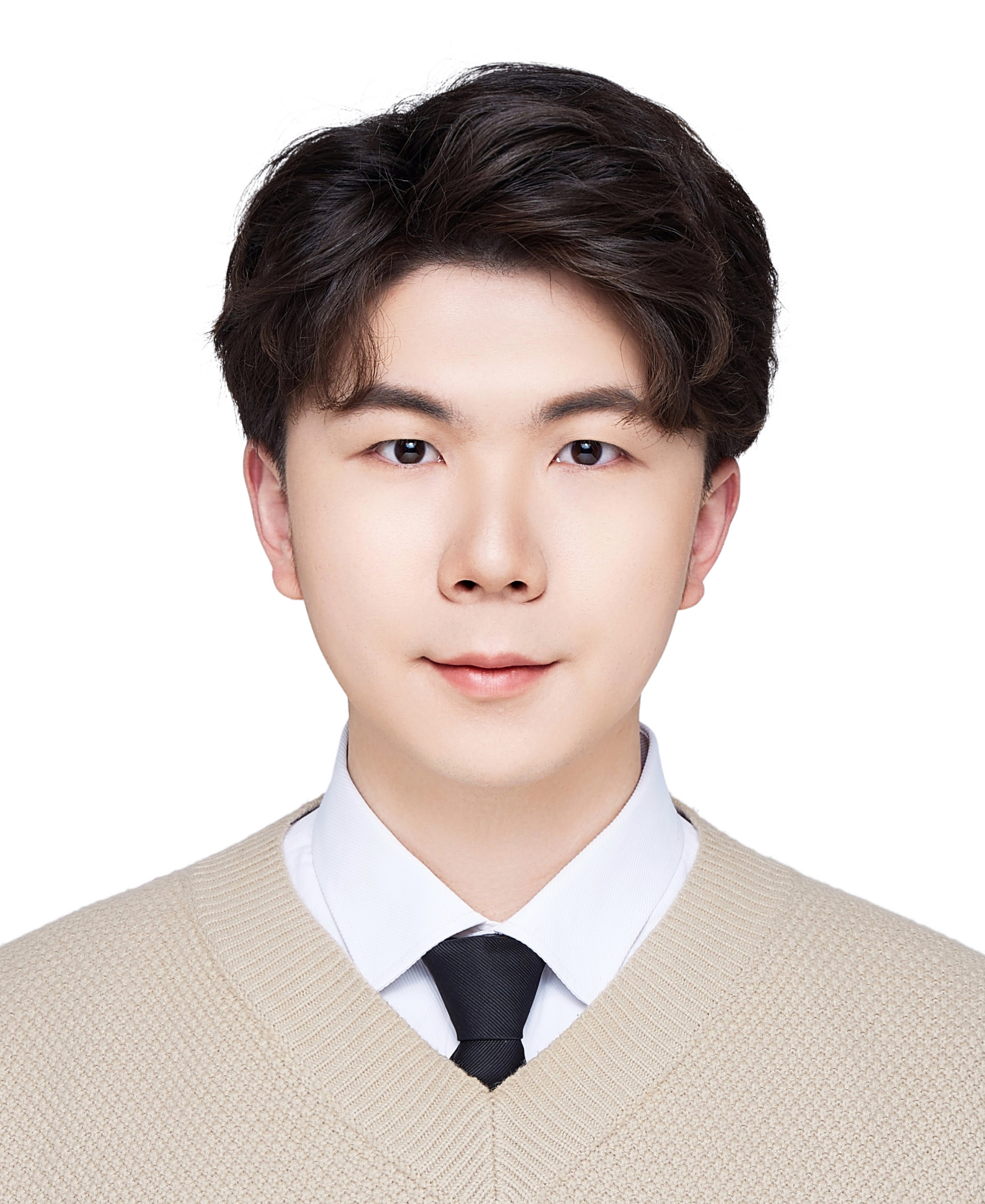}}]{Yifeng Peng} (Student Member, IEEE) received the B.E. degree in Electrical Engineering at University of Electronic Science and Technology of China. He is currently a Ph.D. student in the School of System and Enterprises at Stevens Institute of Technology. His research areas include machine learning in 5G vulnerability detection.
\end{IEEEbiography}

\begin{IEEEbiography}[{\includegraphics[width=1in,height=1.25in, clip,keepaspectratio]{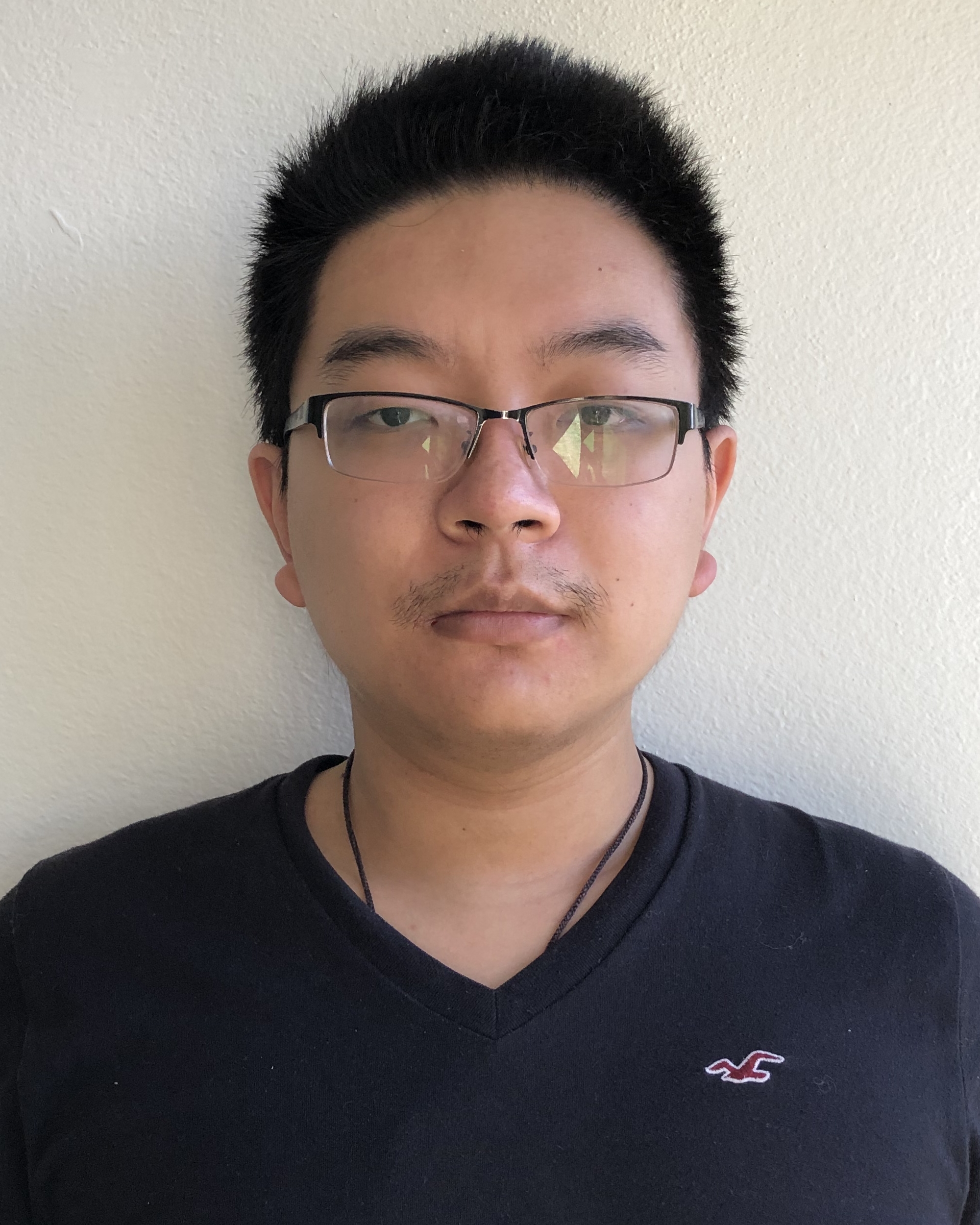}}]{Jingda Yang} (Student Member, IEEE) 
received a B.E. degree in software engineering from Shandong University and an M.Sc. degree in computer science from George Washington University. He is currently a Ph.D. student in the School of System and Enterprises at Stevens Institute of Technology.  His research interests are formal verification and vulnerability detection of the wireless protocol in 5G.
\end{IEEEbiography}

\begin{IEEEbiography}[{\includegraphics[width=1in,height=1.25in, clip,keepaspectratio]{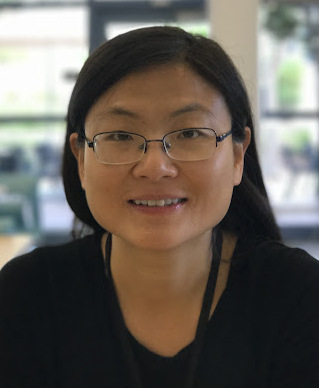}}]{Ying Wang} (Member, IEEE) received the B.E. degree in information engineering at Beijing University of Posts and Telecommunications, M.S. degree in electrical engineering from University of Cincinnati and the Ph.D. degree in electrical engineering from Virginia Polytechnic Institute and State University. She is an associate professor in the School of System and Enterprises at Stevens Institute of Technology. Her research areas include cybersecurity, wireless AI, edge computing, health informatics, and software engineering. 
\end{IEEEbiography}

\vfill

\end{document}